# The M$^3$ project: 2 - Global distributions of mafic mineral abundances on Mars

Riu L., Poulet F., Bibring J.-P., Gondet B.

**Abstract**

A radiative transfer model was used to reproduce several millions of OMEGA (Observatoire pour la Minéralogie, l'Eau, les Glaces et l'Activité) spectra representative of igneous terrains of Mars. This task provided the modal composition and grain sizes at a planetary scale. The lithology can be summarized in five mineral maps at km-scale. We found that the low albedo equatorial regions of the Martian surface (from 60°S to 30°N) are globally dominated by plagioclase with average abundance ~50 vol% and pyroxenes with total averaged abundance close to 40 vol%. An evolution of the LCP/(LCP+HCP) ratio is observed with time at the global scale, suggesting an evolution of the degree of partial melting throughout the Martian eras. Olivine and Martian dust are minor components of the modelled terrains. The olivine distribution is quite different from the other minerals because it is found on localized areas with abundance reaching 20 vol%. A statistical approach, to classify the pixels of the abundances maps, using *k-means* clustering, highlighted seven distinct mineral assemblages on the surface. This classification illustrates that diverse mineralogical units are found in the Noachian and Hesperian terrains, which suggests the presence of various and complex magmatic processes at a global scale during the two oldest eras. The chemical composition was derived from the modal composition maps. The OMEGA-derived chemical composition is quite consistent with several distinctive geochemical characteristics previously considered as fingerprints of the Martian surface. A major discrepancy is in regards to the Fe content that is significantly smaller than soil and rock analyses from GRS and in situ measurements. The discrepancy could be partly explained by the assumptions used for the spectral modelling or could also indicate surface alteration rinds.

## 1-Introduction

The purpose of this study is to provide global distributions of Martian surface materials found in mafic regions. This is distinct from global previous investigations based on OMEGA data set (Bibring et al. (2005), Poulet et al. (2007), Ody et al. (2012)), which determine mineral distribution based on spectral signatures but did not provide global estimates of abundance. This work continues the previous efforts described in the companion paper (Riu et al., 2019) by focusing on the derivation of the modal mineralogy of mafic terrains as defined for the $M^3$ (Modal Mineralogy of Mars) project. The compositions and distributions of mafic materials on the surface should highlight unique conditions and processes that have occurred during Mars' history. As future space exploration will conduct sample return with quantitative analyses of the rocks, future analyses will use the dataset displayed here to provide global context for those analyses.

Previous quantitative investigations at global scale were done by the Gamma Ray Spectrometer (GRS on board Mars Odyssey) and the Mars Global Surveyor (MGS) Thermal Emission Spectrometer (TES) instruments. Although the derivation of the mineralogy from GRS elemental composition is not straightforward, the TES data set was conversely extensively used to derive the modal mineralogy of spectrally distinct Martian low-albedo regions and to identify spatial trends in mineralogical assemblages at a global scale (Rogers and Hamilton, 2015 and references therein). They have confirmed previous works by Bandfield et al. (2000) and Christensen et al. (2000) that plagioclase, pyroxenes and high-silica phases dominate the global surface composition. The unambiguous detection of low-calcium pyroxene referred to as LCP hereafter (> detection limit of ~15 wt%) with TES was established by Rogers and Christensen (2007). Relatively higher abundance was especially spotted in northern Acidalia and Solis Planum, though the total pyroxene (LCP + HCP) abundance was relatively low in these regions. The resulting maps of pyroxenes predict to find more low-calcium pyroxene (LCP) than high-calcium pyroxene (referred to HCP hereafter) on the overall surface, with lowest LCP values found in the Syrtis Major edifice. Olivine is commonly located in the southern highlands with abundance ~10 to 20 wt% and appears to be more common near the north/south dichotomy (Koeppen and Hamilton, 2008). Chemical analysis with GRS reveals that the surface is rather homogeneous in terms of chemical composition compared to Earth. Silicon and iron, which are dominant, are homogeneously widespread with higher abundances in the northern lowlands for the later (Boynton et al., (2007), Taylor et al., (2010)). Minor components that vary spatially (K, Th, Cl, H) have been also detected and mapped providing major constraints on the evolution of the mantle petrology (Taylor et al., 2010; Baratoux et al., 2013).

Local scale analyses of the rocks on the floor of Gusev Crater carried out by the Spirit Rover have highlighted the presence of abundant basalt with olivine phenocrysts. These picritic basalts have a composition close to that of the olivine-phyric Shergottites (McSween et al., 2006). Additionally, the first alkali-rich lithologies were characterized in the Noachian terrains of Columbia Hills (Ruff et al., 2006). As for the MSL Curiosity rover, it has discovered, in the Hummocky plains, several felsic rocks covering a large range of composition from alkali basalt to trachyte (Sautter et al., 2016). These observations were used to suggest the possible formation of a continental crust on Mars in its early stages that was not envisioned from orbit and previous *in situ* measurements (Sautter et al., 2015; Sautter et al., 2016).

Globally, combination of those numerous *in situ* and orbital analyses of different scales and techniques have led to the conclusion that the global composition of Martian crust is best explained by tholeiitic basalts (McSween et al., 2009a; Taylor et al., 2010; Grott et al., 2013; Sautter et al., 2016).

Interpreting NIR data of mafic regions from OMEGA spectra was previously conducted using a quantitative retrieval of mineral abundances from the modelling of spectra of selected terrains (Poulet & Erard, 2004; Poulet et al. 2009a, 2009b). These studies of mineral detection and composition modelling were restricted to selected terrains and based on the Shkuratov radiative transfer theory (Shkuratov et al., 1999). The derived modal mineralogy was similar to basaltic/tholeiitic composition dominated by plagioclase, high-calcium pyroxene and low-calcium pyroxene. The LCP/(HCP+LCP) ratio indicates that HCP is always dominant over LCP, with the lowest relative LCP values in Syrtis Major and largest values between Valles Marineris and Argyre impact basin (Poulet et al., 2009a). Olivine has also been found in more localized areas with strong variable abundances depending on the spots (from less than 10% up to 40%).

This work is an extension of these previous regional analyses. It benefits from three recent developments: 1) the construction of a global 3D reflectance map (companion paper 1, Riu et al., 2019) that allows the extraction of any spectrum of interest over the entire surface of Mars of latitudes ranged between 60°S and 60°N. 2) The optimization of the radiative transfer code and minimization procedure diminishing considerably the computing time of the fitting procedure. This permits numerous global simulations to test the sensitivity of the fits to many parameters required to evaluate the quality and the uncertainties of the final abundance maps. 3) The cheaper and more performant computing processor units (CPU) increasing the number of CPUs and making possible the launch of tens of simulations at the same time.

Before applying systematically the radiative transfer model based on the Shkuratov theory (Shkuratov et al., 1999), different model tests werecarried out to evaluate the sensitivity of various parameters to the derived abundances (Section 2). The resulting compositional (either mineralogical or chemical) maps are presented and discussed in sections 3, 4 and 5 respectively. Analyses were restricted to 60°S to 30°N in latitude (section 2.1) and to locations presenting pyroxene signatures in the NIR, which mostly correspond to low albedo terrains. The overall comparisons with previous remote sensing measurements from the other global studies mentioned above and from *in* situ measurements are also discussed.

## 2- Derivation of modal mineralogy

### 2.1. Data selection, modelling method and building of the mineral maps

As shown with previous studies at regional scales (Poulet et al., 2009a; 2009b), orbital NIR reflectance spectra of igneous terrains can be modelled to extract modal composition of regolith-like surface. The selection of the spectra to be modelled is based here on the OMEGA pyroxene band depth criterion (Ody et al. 2012; Riu et al., 2019, companion paper 1) applied to the OMEGA-based global spectral cube. Only the spectra having a pyroxene spectral index (e.g. band depth) > 0.02 are considered for further analyses, which corresponds to ~12 million of spectra between -60° to +60° in latitude. Regions exhibiting olivine signature alone, such as Nili Fossae are not taken into account in our study. The pyroxene distribution in terms of band depth, prior to the modelling, is shown in Figure 1.

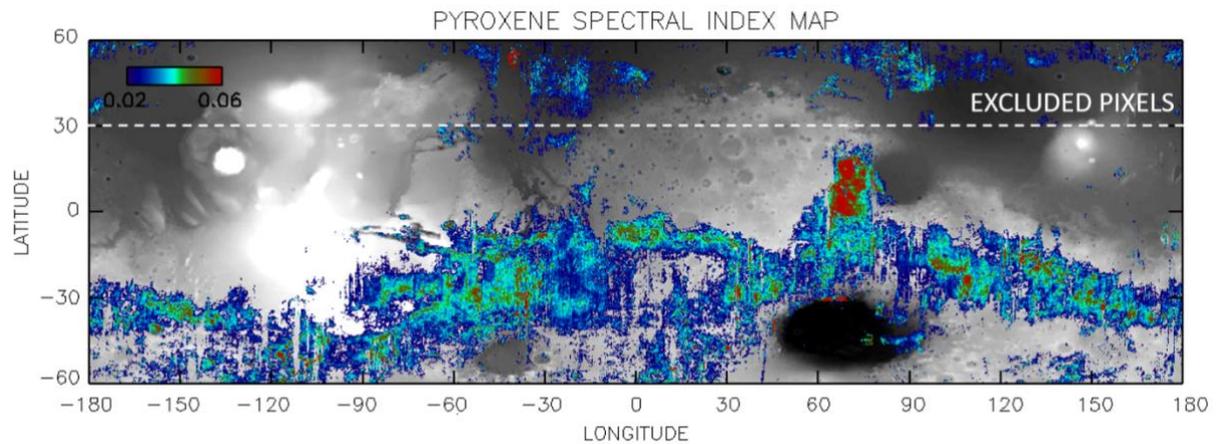

Figure 1 – Pyroxene index map derived from the 3D global reflectance map of Mars (Riu et al., 2019) using spectral index from Poulet et al. (2007) and Ody et al. (2012). For modelling purpose, the threshold is fixed at 2% and the northern plains are excluded from our sample.

The modelling method is based on the Shkuratov radiative transfer theory, which has been adapted to deal with basaltic composition (Poulet and Erard, 2004). The application of this theory is restricted to powderer-like surface, which limits its application to the entire surface of Mars. Specifically, the northern plains spectra exhibit unique spectral features in the near-infrared wavelengths with very weak pyroxene signatures overlapping a strong blue slope (Mustard et al. 2005; Poulet et al. 2007a). These spectral properties are best explained by a dust/silica coating (Minitti et al., 2007, Poulet et al. 2007b, Salvatore et al., 2013), or leached rinds of basaltic glass (Horgan et al., 2017). As the current model is not able to simulate the radiative transfer for such physical properties, we removed from further modelling the latitudes from 30°N to 60°N (Figure 1). This selection still gives a total of 10.3 million of spectra to be modelled in the 60°S to 30°N in latitude region.

To reproduce OMEGA reflectance spectra, it has been shown that five end-members with varying abundance and grain size shall be sufficient (Poulet et al., 2009a): augite for the high-calcium pyroxene end-member, pigeonite for the low-calcium pyroxene end-member (that will be respectively referred to as HCP and LCP in the following), forsterite for the olivine (that will be also referred to Mg-rich olivine or olivine in the following), palagonite (JSC Mars-1, (Allen et al., 1997) for the dust and labradorite for the neutral plagioclase end-member (see Figure **2**2). The model thus enables the quantification of mineral phases with shallow absorption features (section 3), that are hardly detected with spectral indexes method. The selection of the mineral endmembers has been discussed in detail in Poulet et al. (2009a) and validated for a variety of terrains (Poulet et al. 2009b). As a reminder, this mineral selection is a tradeoff between initial guess resulting from OMEGA/CRISM spectral signatures, previous studies (mainly TES and *in situ*), availability of the optical constants, number of free parameters and computing time. A low-albedo component is added as small inclusions (< 10% in volume) in the plagioclase to lower the average reflectance and reduce spectral contrast as explained in Poulet et al., (2009a). Magnetite is used to account for this behavior. The use of palagonite as the dust end-member in the low albedo terrains is a new end-member in comparison of the previous studies; this choice will be discussed in section 2.3. An additional end-member corresponding to basaltic glass is present on Figure 2. The modelling with basaltic glass will be discussed in section 2.8.

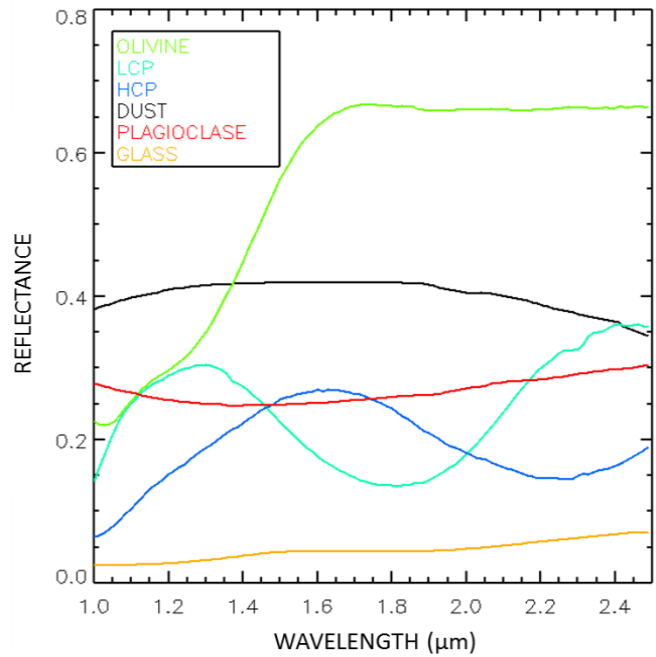

Figure 2 - Reflectance spectra of the different end-members used in the modelling. The olivine, LCP, HCP and plagioclase spectra come from Poulet et al. (2009a). The glass spectrum is extracted from Cannon et al., (2017). The palagonite spectrum is in black. The grain size used for each end-member spectrum plotted on Figure 2 is extracted from Table 1.

|  | Palagonite (Dust) | Augite (HCP) | Pigeonite (LCP) | Labradorite (Plagioclase) | Forsterite (Olivine) | **Basaltic glass** |
|---|---|---|---|---|---|---|
| Grain size (µm) | 5 | 100 | 100 | 150 | 100 | 100 |
| Abundance (vol %) | 5 | 25 | 25 | 35 | 10 | 15 |

Table 1 - Starting conditions of the modelling for grain size and abundance of each end-member.

For each modelled spectrum, the model derives an abundance (in vol%) and a grain size (in µm) for all end-members except for the magnetite present as inclusion where only the abundance is reported. The olivine and dust end-member grain sizes are fixed as explained in

(Poulet et al., 2009a). The initial conditions that are representative of the average expected abundances and grain sizes are reported in Table 1.

A spectral slope is also included as a parameter to account for possible remaining aerosol contribution in spectra and photometric effects. For each OMEGA spectrum, the model thus provides a synthetic spectrum with 12 parameters (abundances in vol %, grain sizes in µm and slope) including 9 as free parameters. To obtain the best fit, a downhill simplex method for multidimensional minimization routine is used in order to minimize the RMS (Residual Mean Square) between OMEGA and modelled spectrum. The resulting RMS is then used as a parameter to quantify the quality of the fits, once the best-fit is reached. The acceptance threshold on the RMS is 0.003 as established in Poulet et al., (2009a). The minimum RMS depends on the optimization pathway, which in turn depends on the initial grain sizes and abundances used in the model. Thus, though all pixels are required to be fit with an RMS value < 0.003 (that only represents the upper limit to which a fit is deemed acceptable), it is possible that the minimum RMS value (below 0.003) can vary with initial conditions. The effect of initial conditions on final modelled abundances is evaluated in Section 2.5. This approach was already used in Poulet et al., (2009b), but it will be extended to the full selected data set described here. When the fitting procedure reaches a minimum RMS value, hence a best-fit, the abundances can be retrieve for each pixel. The sum of the abundances for every spectra is 100vol%. Additionally, individual abundances cannot be negative but they are allowed to reach 0vol%. The complete scheme from spectral data to modal composition is summarized in Figure 3. As an addition to the modal composition, which is of interest here, the chemical composition can eventually be calculated from the mineralogy using the end-members oxides composition.

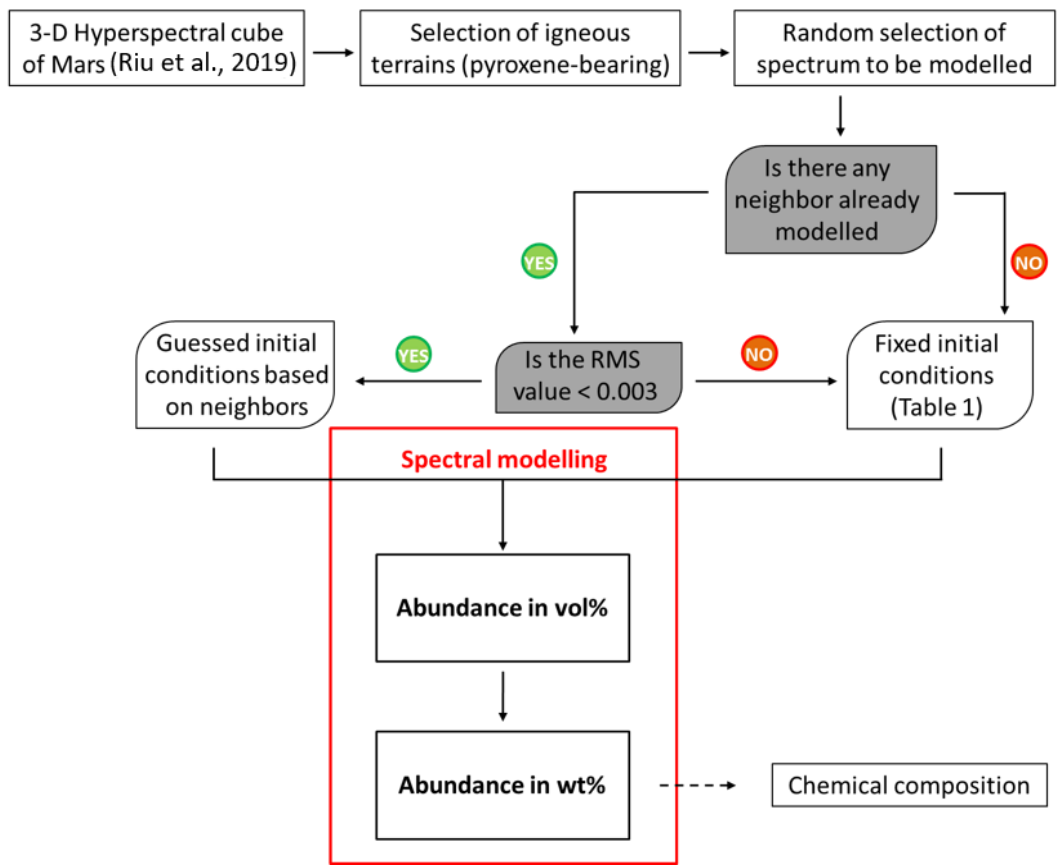

Figure 3 – Organigram describing the pipeline of modelling from spectral data to modal mineralogy and eventually chemical composition.

The global cube was divided into 20 quadrangles allowing the model to be launched over these regions in parallel using several tens of CPU; in the following sections, different regions may be referred to as MC# (where # is the corresponding Martian quadrangle). As the fitting is an iterative procedure, the convergence towards the best fit for a given spectrum is faster if the initial conditions for abundances are close to the resulting one. The initial conditions regarding the abundances were thus fixed to initial guessed values when possible (Figure 3). Specifically, the filling of the map is carried out randomly pixel by pixel and the corresponding initial conditions are, if possible, calculated from the modelling result of its closest neighbors. A pixel is thus selected randomly and three possibilities might occur: (1) none of its neighbors is already modelled; in this case, the initial conditions used are the ones established in Table 1.

(2) One of the neighbors is already modelled with a resulting RMS < 0.3 %; then, the initial conditions (both abundances and grain sizes) are those derived from the best fit of the neighbor. (3) Several neighboring pixels are fitted with a resulting RMS < 0.3%; then the initial conditions correspond to the averaged obtained derived abundances and grain sizes of several neighbors. This map filling process is based on the assumption that the abundances should not vary significantly from one pixel to its closest neighbors. This procedure leads to less iterations in the fitting procedure, and thus decreases the computing time of the minimization process (Figure 4a). It was also checked that not only the fit procedure was faster but also the resulting fits were improved for most of the spectra when the initial conditions used were the one selected from the result of neighbors, as illustrated on Figure 4 (b).

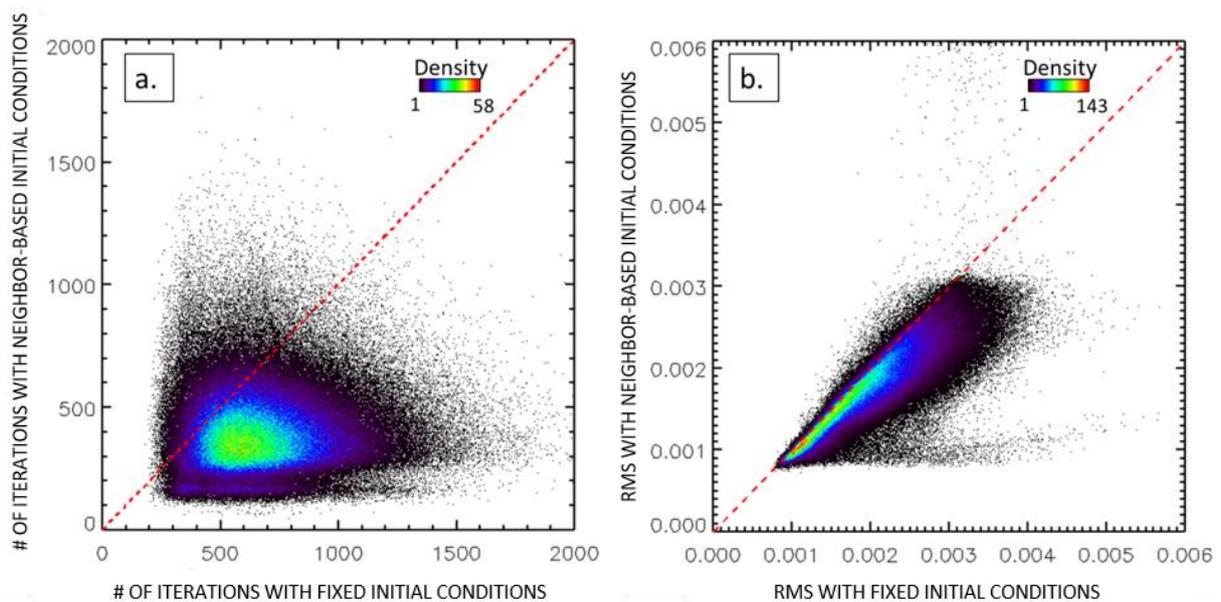

Figure 4 - (a) Number of iterations of the fitting procedure using guessed initial conditions with respect to the number of iterations required with fixed initial conditions (Table 1). (b) Final RMS values with guessed initial conditions with respect to the final RMS values with fixed initial conditions. This test was carried out on ~350000 spectra over the MC13 quadrangle.

The quality of the fit could depend on the seed used for the random map-filling process. In order to prevent this bias, several seeds were tested and combined to build the final map. As shown in Figure 55, the percentage of acceptable fits increases with the number of maps (i.e. seeds) that are merged. Each of the maps was built using different random seeds, so that the probability for each pixel to have been modelled with initial guessed conditions increased. This method enabled the launching of the model on all available computer processing units at the same time, which reduced the total computing time.

The final modelled maps are a combination of 10 simulations where the final set of fitting parameters for each pixel is provided by the best fit of the 10 fits. As stated before, the computing time is a major issue in this modelling process so the number of seeds tested has been limited. The limit of 10 simulations was established as a trade-off between computing time and the asymptotic evolution of good fits as a function of the number of maps (Figure 5).

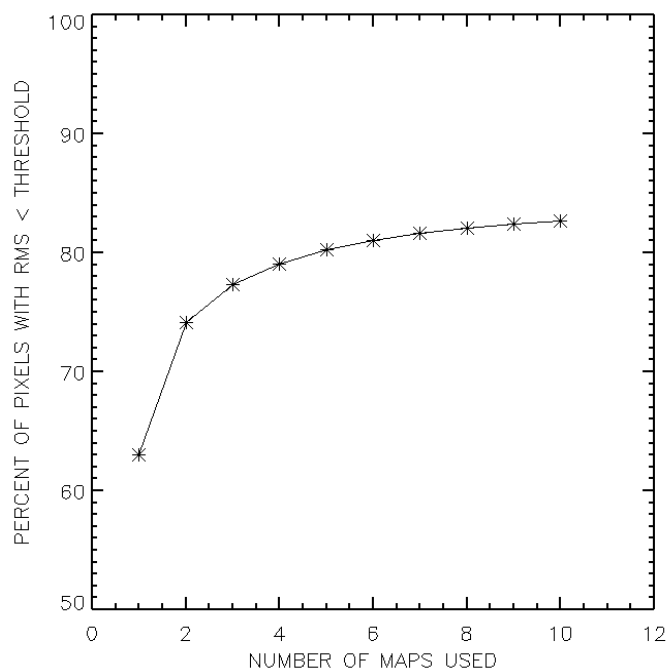

Figure 5 - Evolution of the percent of pixels with RMS < threshold (0.0030) versus the number of simulations (i.e. seeds) used to derive the global abundance maps.

The median value of the RMS is 0.0021, while 17% of spectra (and thus pixels) have RMS > 0.003 (Figure 6). Those rejected modelled spectra were analyzed and classified in order to understand why the reflectance spectra were not successfully modelled. They are mainly located in the southern hemisphere at latitudes below 30°S. The major causes were found and listed here with increasing number of assigned spectra: weak spurious instrumental artifacts that were not taken into account in the data reduction scheme of OMEGA, presence of weak water ice signatures (likely under the form of clouds) unfiltered by the water ice exclusion criterion (see section 2.1 of companion paper 1), high albedo (> 0.25) indicating the possible presence of large amount of dust and/or additional unknown end-member, weak SNR in part as the result of bad atmospheric corrections and observational conditions. In the following, those pixels were removed from the global mapping leading to a sample of ~8.1 million modelled spectra with RMS < 0.003.

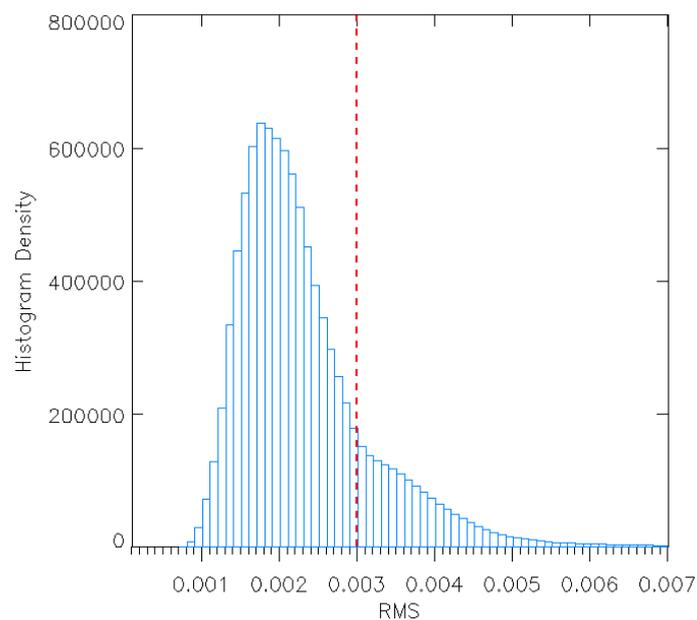

Figure 6 - RMS histogram distribution. The red dotted line represents the RMS threshold.

2.2. Influence of the building method

As the final maps are obtained from the combination of several maps, it is important to check if this merging method has an influence on the resulting abundances and grain sizes. The inspection consists in plotting the average abundances of each mineral as a function of the number of simulations used for the merging. The variations for each mineral are shown on Figure **7** 7.

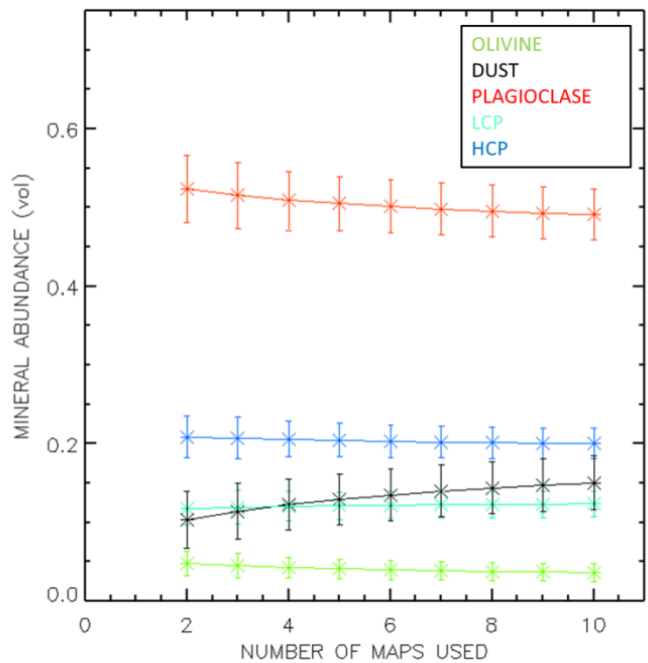

Figure 7 - Evolution of mineral abundances for each global map with respect to the number of simulations (i.e seeds) used for deriving the final abundances. The error bars represent by the standard deviation of the abundances.

The averaged abundance of plagioclase starts at 53 vol% to reach a final average abundance of 50±4 vol%. The high-calcium pyroxene end-member is the second most abundant mineral on the surface with an average of 20±3 vol%. Its abundance varies very little with the number of simulation. The abundances of the dust and low-calcium pyroxene are of the same order as the high-calcium pyroxene and are respectively 15±4 vol% and 13±2 vol%. We however note that the dust abundance significantly increases with the number of simulations, in a proportion opposite to the plagioclase one. This indicates that the abundances of these two endmembers that are known to have only weak spectral signatures in the near-infrared are correlated. The sum of the two abundances is actually constant with the number of simulations. Among the selected end-members and on the overall sampled surface, the less abundant mineral is the olivine with an average abundance of 4.0±1.5 vol%. It will be however shown in Figure 16 (c) the abundance of olivine can reach 20% in more localized areas.

The variation of the abundance with the number of simulations for individual spectra is also of the same order than the standard deviation on the global averaged abundance as shown in Figure 8.

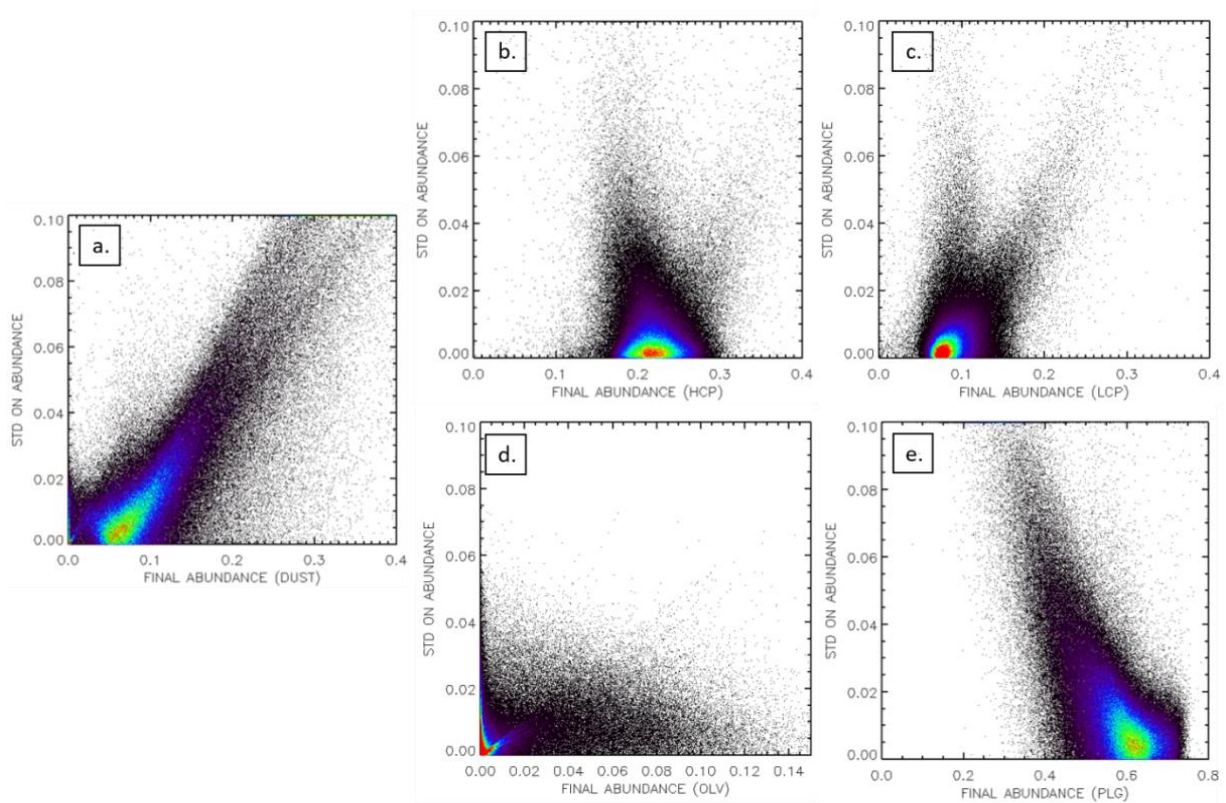

Figure 8 - Density plots of the final abundances for each end-member with respect to the STD on abundances for individual spectra. The STD of each spectrum is calculated from the 10 maps. (a) Dust. (b) HCP. (c) LCP. (d) Olivine. (e) Plagioclase. The computation of the STD was made on the MC13 quadrangle.

Overall, the variations are relatively small and for the majority of pixels the standard deviation is still below the final estimation of error (Table 2), leading us to be confident on the final values obtained for the abundances of each mineral. These abundances and thus the spatial variations on the surface of Mars will be discussed in greater details in section 3.

2.3. Influence of the choice of dust end-member

In previous regional studies, the same end-members were used and validated in restricted regions (Poulet et al. 2009b). Among them, the dust end-member was very specific in the sense that it was computed from OMEGA measurements. For the global scale study, additional tests were carried out on the choice of this end-member. Two Martian dust analogs were included in order to quantify and prevent possible bias from the OMEGA-derived dust spectrum: palagonite (Bell et al. (1993), Allen et al., 1997, Morris et al., 2001) and nanophase hematite (Morris et al., 1997). These two end-members were considered as good analogues to explain the visible spectral properties of the dusty regions of Mars. An additional test was also implemented by excluding any dust phase from the list of initial end-member. The percentage of fits with satisfying RMS as a function of dust end-member is shown in Figure 9. It shows that more spectra are fitted with RMS values below the threshold when palagonite is used (Figure 9a). As a result this end-member will be selected for the global modelling to account for the Martian dust analog of the low albedo regions modelled in our study. However, it should be noted that for dustier regions, which are excluded from our sample and thus not modelled here, another dust analog may be more representative.

The quality of the fits is an important parameter but it is also essential to see how the abundances vary as a function of the dust end-member. The same trend for each set of simulation is observed: dominant mineral is plagioclase (>50%), the less abundant phase in the mixture being olivine (~5%), the three other are in-between with average abundances ranging from 10% to 25% in the dark modelled regions. The simulation with palagonite is the only one with average dust abundance larger than the average low-calcium pyroxene abundance. It appears that the results obtained with the simulation "No dust" differs most from the other simulations (Figure 9b).

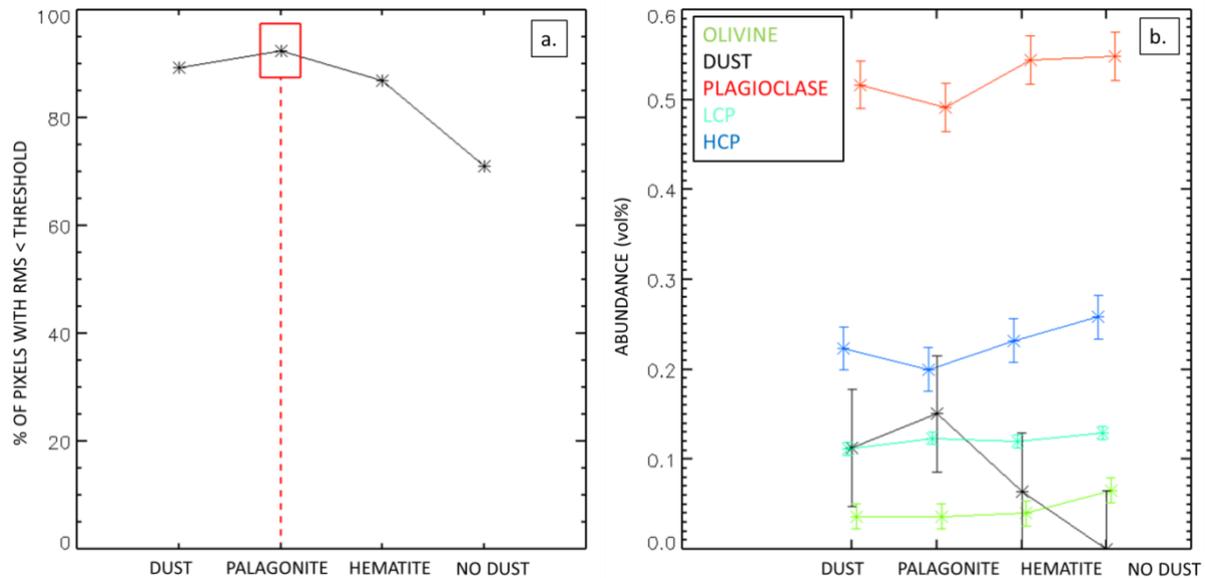

Figure 9 - (a) Evolution of the percentage of good fits (RMS ≤ 0.003) with respect to the end-member used for the dust analog. The case named "Dust" defines the dust end-member originally used for the regional studies (Poulet et al. 2009b). (b) Evolution of the average abundance for each end-member with respect to the dust end-member used for the global simulation. The error bars correspond to the standard deviation of the average values.

2.4. Influence of the grain size

Previous works indicated that the grain size is poorly constrained (Poulet and Erard 2004; Poulet et al. 2009a). Because the spectral signatures (depth and width) depend on the grain size, we here explore the influence of the grain size on the resulting abundance. A set of simulations was performed with two different initial grain sizes for the LCP and HCP end-members: 100 (nominal starting condition) and 200 µm. The comparison is restricted to the fits with RMS ≤ 0.003. A second criterion on the absolute RMS difference between the two sets of modelling is also applied in order to keep only spectra with similar fit quality. This criterion is the following: two modelled spectra of the same pixel are compared if the final RMS values of two fits differ by less than 1%. For those spectra respecting these two conditions (which corresponds to about 16% of the entire sample), we compared the resulting abundances and grain size (Figure 10). As the olivine and dust grain sizes are fixed, the influence on the initial grain size is mapped for HCP, LCP and plagioclase only.

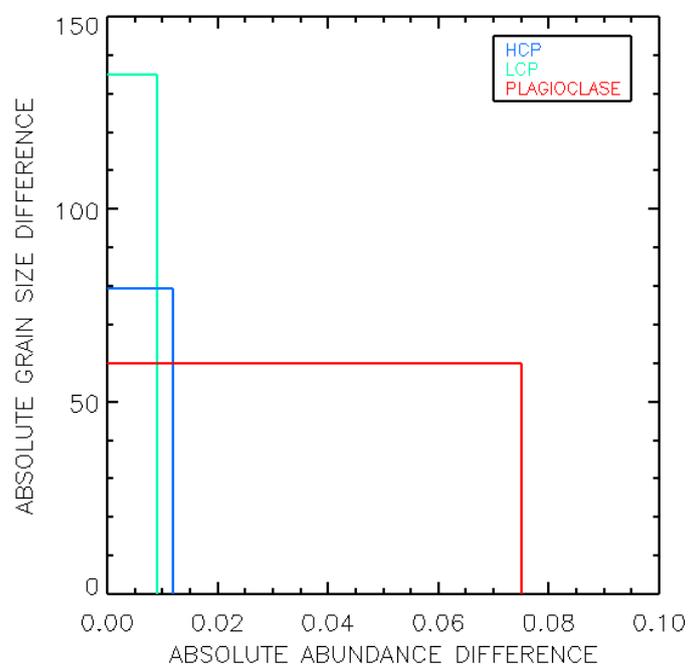

Figure 10 - Absolute abundance difference (in vol%) versus absolute grain size difference (in µm) between two sets of modelling starting respectively with 100 and 200 µm grain size for the pyroxenes. For clarity purpose, the differences are represented by boxes that contain more than 99% of the fits.

For the selected spectra, a wide range of grain sizes (with a maximum difference of about 135 µm for the same spectrum for LCP) is observed; the abundances are relatively less affected when changing the initial conditions on the grain size as the resulting differences are of the order of ±1 vol% for LCP and HCP and ± 7.5 vol% for plagioclase. Based on this test, it looks that the grain size initial condition plays a minor role on the resulting abundances.

2.5. Sensitivity to the initial conditions

The quality criterion is based on the RMS value. This is a convenient and usual way to assess the quality of the fit, however there is no unique solution once the RMS < 0.3 % and the optimization pathway will depend on the initial conditions. It is thus important to evaluate for a given spectrum the number of fits respecting RMS < 0.3% starting with different initial conditions and their resulting abundances and grain sizes. This test requires performing the modelling of the same spectrum by varying the initial conditions. This test was restricted to a limited number of spectra representative of the spectral diversity of the low albedo regions. Specifically, the method consisted in launching the modelling 1000 times on a given spectrum with varying initial conditions selected randomly. Figure 11 illustrates the accepted fits for 5 spectra tested with this method. More than 90% of 1000 simulations provided a best fit with a RMS value < 0.3%.

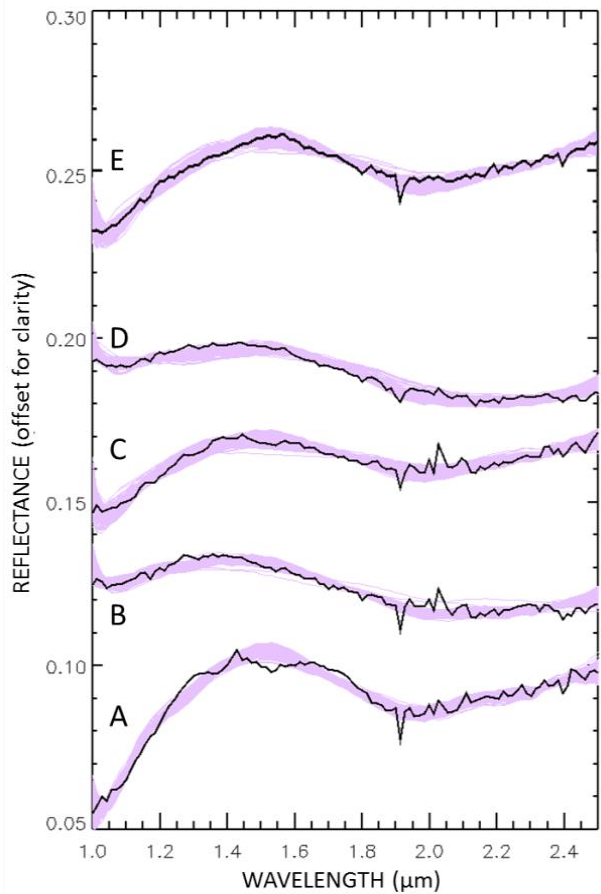

Figure 11 – Five OMEGA spectra (black) compared to their best fits (pink) among the 1000 simulations launched with random initial conditions.

For the 5 spectra, the standard deviation for the abundances among the acceptable fits (RMS <0.003) is on average 7.8 vol% for the dust analog, 4.6 vol% for HCP, 4.0 vol% for LCP, 2.6 vol% for olivine and 8.2 vol% for plagioclase. The resulting standard deviations on the abundances are thus relatively low and proportional to the average abundances. For grain size, the standard deviation is of 100 µm for all end-members. Those additional variations will have to be taken into account when computing the overall uncertainties on the abundances on the final maps.

2.6. Influence of the minimization routine

Due to the nonlinear formulation of the Shkuratov theory, the inversion problem has to be solved using an iterative approach to minimize the RMS value. As shown in Figure 7, the RMS distribution is centered on ~0.002, which means that the majority of the fitted spectra has RMS lower than the threshold. But thousands of possible mixtures are tested before reaching a final solution, corresponding to the best-fit (RMS ≤ 0.003), for a given spectrum. In order to illustrate the range of values considered in the fitting effort and provide an estimate of the uncertainty on the final results resulting from the iterations, we followed the evolution of the abundances and grain sizes during the iterative pathway from when the threshold value is reached (RMS = 0.003) up to its final RMS value < 0.003 that ultimately corresponds to the best-fit (Figure 12). For this particular example, the best-fit is found for a RMS value of 0.001. Note that the simulations were launched without guessed initial conditions.

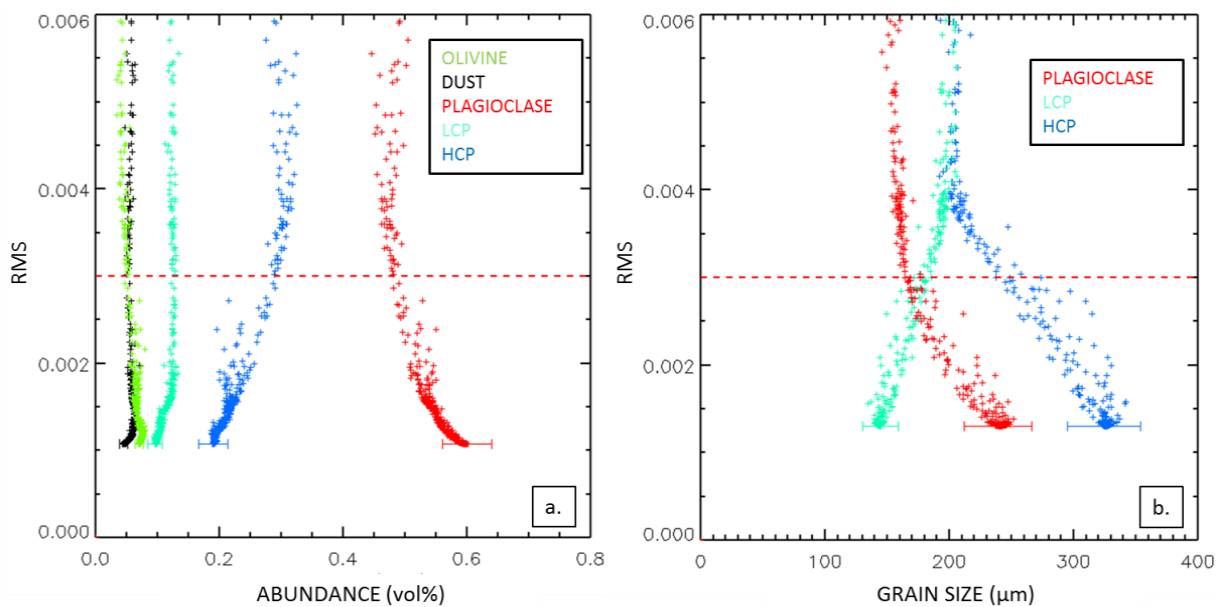

Figure 12 – (a): Evolution of the mineral abundances with respect to the RMS during the minimization for one pixel in MC13 (Syrtis) region. (b): Evolution of the grain size (for each mineral with variable grain size) with respect to the RMS. The error bars represent the abundance (left) and grain size (right) standard deviation from RMS=0.003 to final RMS value. Red dotted line indicates the RMS threshold.

As the computer recording of abundances and grain sizes at every step of the minimization is extremely memory-consuming, we decided to follow this evolution for one of the quadrangles only (MC13), which corresponds to ~400000 spectra. As long as an improvement of the fit is possible, the minimization routine indeed continues to explore the parameter space of abundance and grain size leading to larger variations. The exploration depends on the endmember, but the derived standard deviations are small: < 3 vol% on average for the abundances and < 15 µm for the grain sizes. These uncertainties are thus rather small, especially for the grain sizes (Figure 13). We are thus confident on the "consensus" of the solutions in terms of modal mineralogy. The resulting average standard deviation will be used as uncertainties that will be taken into account in the final estimate on the uncertainties of the modal mineralogy.

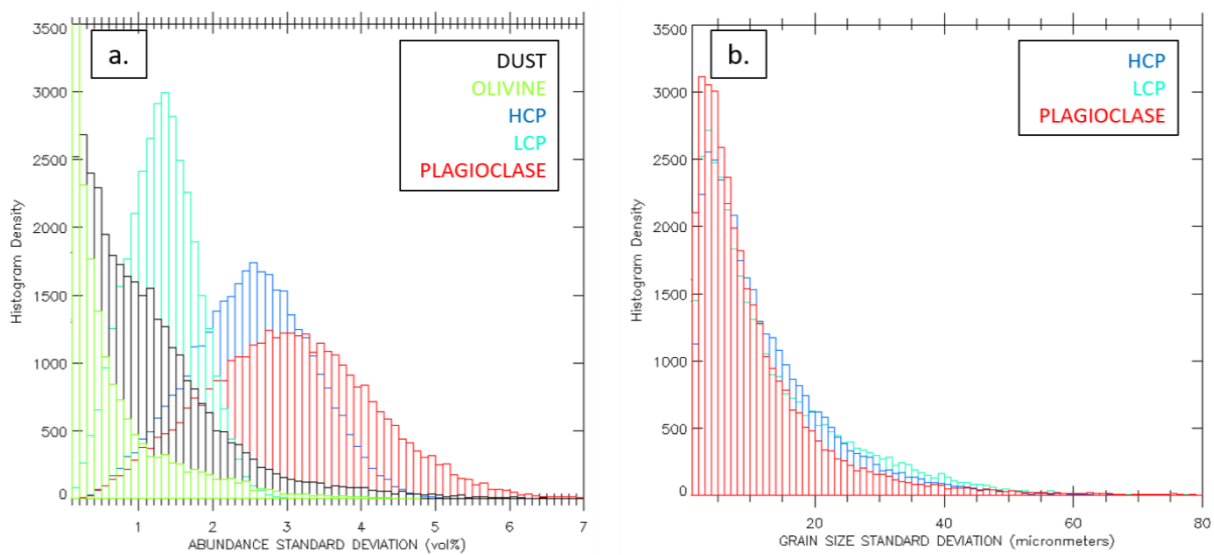

Figure 13 – (a): Histograms of abundance standard deviation due to the exploration of the values by the minimization routine for all pixels on MC13 (Syrtis) region. (b): Same as (a) for the grain size.

2.7. Errors from OMEGA radiometric calibration

The strength of the modelling approach is to use all the spectral information contained in a spectrum, namely spectral signatures, spectral slope but also reflectance level (Poulet and Erard, 2004). Therefore, the systematic error on the reflectance level of any OMEGA spectrum may have an impact on the resulting values. From the calibration campaign of the OMEGA instrument, it has been estimated that the absolute radiometric performance of OMEGA is of the order of 20% (Bonello et al., 2005). The use of the 3-D global hyperspectral map of Mars also contributes to the error on the overall reflectance. As described in the companion paper 1 (Riu et al., 2019), this error is on average about 3%. To investigate the sensitivity to these errors, we then modelled ~150000 spectra from various locations of Mars while varying their overall reflectance level by steps of 5% in the ± 25% range. The derived variations on the abundances and grain sizes were found to be 4 vol% and 9 µm for the plagioclase, 0.8 vol% and 8 µm for the HCP, 0.55 vol% and 7 µm for the LCP. For the palagonite and the olivine, we found 2.6 vol% and 0.15 vol% errors respectively.

## 2.8. Investigation on the presence of a basaltic glass component

Silicate glasses have been proposed to be a component of the regolith of Mars as the results of volcanic and impact processes (Bandfield et al., 2000, Hamilton et al., 2001, McSween et al., 2003). Moreover, the TES-based global mineralogy exibits a significant amount of high silica phases, which could be spectrally analogous in the NIR to such silicate glass. One of the most evident detections was reported in the Hargraves crater situated near Nili Fossae trough (Cannon and Mustard, 2015). However, the glass phases are challenging to detect in the NIR because they have broad and weak absorption features potentially difficult to be extracted when mixed with other minerals such as olivine and pyroxene. As discussed in Poulet and Erard (2004), non-linear mixing has the capability to highlight spectrally featureless phase in remotely sensed NIR data. Therefore, additional sets of simulation with a synthetized analog glass as a $6^{th}$ end-member are here carried out in order to test the presence of this phase.

The glass spectrum used is a synthetized glass representative of the type of basaltic glass that can be found at Gusev Crater (Cannon et al., 2017; Figure 2). It is characterized by significant red slope and absorption feature due to $Fe^{2+}$ crystal field centered on 1.09 µm. The initial conditions for this end-member are the following: abundance of 10 vol% at the expense of the initial plagioclase abundance and grain size of 100 µm. As previously, the final global maps are built from 10 seeds, and the filling process is the one based on the guessed initial conditions.

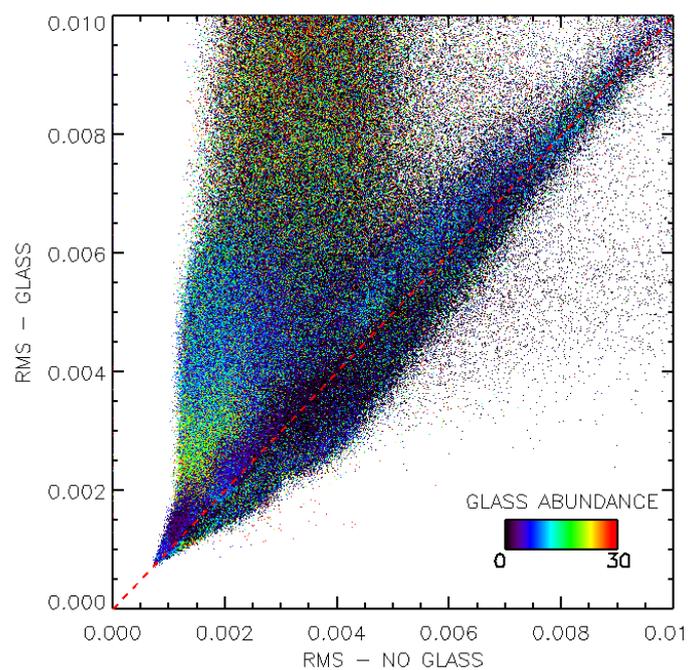

Figure 14 – Comparison of RMS values for the simulation without basaltic glass (simulation "No Glass") and the simulation with basaltic glass (simulation "Glass"). The color scale corresponds to modelled glass abundance. The density of data points is not represented here but 75% of the data points are found below 0.003 for both RMS.

The fits are almost systematically better for the simulation with no basaltic glass (Figure 14). We also observe that the RMS gets better as the glass abundance decreases. However, 75 % of the pixels are found below the 0.003 RMS threshold for both sets of simulations with really similar RMS values. There is also a non-negligible percentage (10 %) of spectra that present a better RMS when including basaltic glass as an end-member. For these pixels, the abundance of glass is however very small, < 5 vol% in average.

To further investigate sensitivity of the modal mineralogy to this component, we select a sample of pixels by retaining all pixels with RMS ≤ 0.003 for both simulations. This additional processing step reduces the sample to 7.6 million pixels. Glass is found roughly homogeneously distributed on the surface with an average abundance of 5 vol%. Pixels (~100000) with large glass abundance of ~30 vol% look also randomly distributed over the surface. The addition of glass does not impact the global spatial distributions of other minerals except for the olivine. We attribute this correlation to the presence of the $Fe^{2+}$ band in the glass spectrum that looks similar to the olivine. We thus explore if some deposits previously recognized as olivine-bearing could be modelled with glass only.

We chose the region of Valles Marineris (MC18) where some spots are well modelled with glass only. Figure 15 shows a comparison of the geographic variabilities of glass and olivine resulting from both simulations. A large olivine deposit located in the center of the scene is replaced by a large glass deposit when basaltic glass is introduced in the simulation. There is also a deposit in the north of this region where the olivine concentration has significantly decreased at the expense of basaltic glass. These observations question the discrimination between those two end-members and the confidence on the detection and identification of olivine-bearing terrains performed in the NIR. However, the quality of the fits (Figure 14) tends to show that olivine is the best end-member to reproduce OMEGA data.

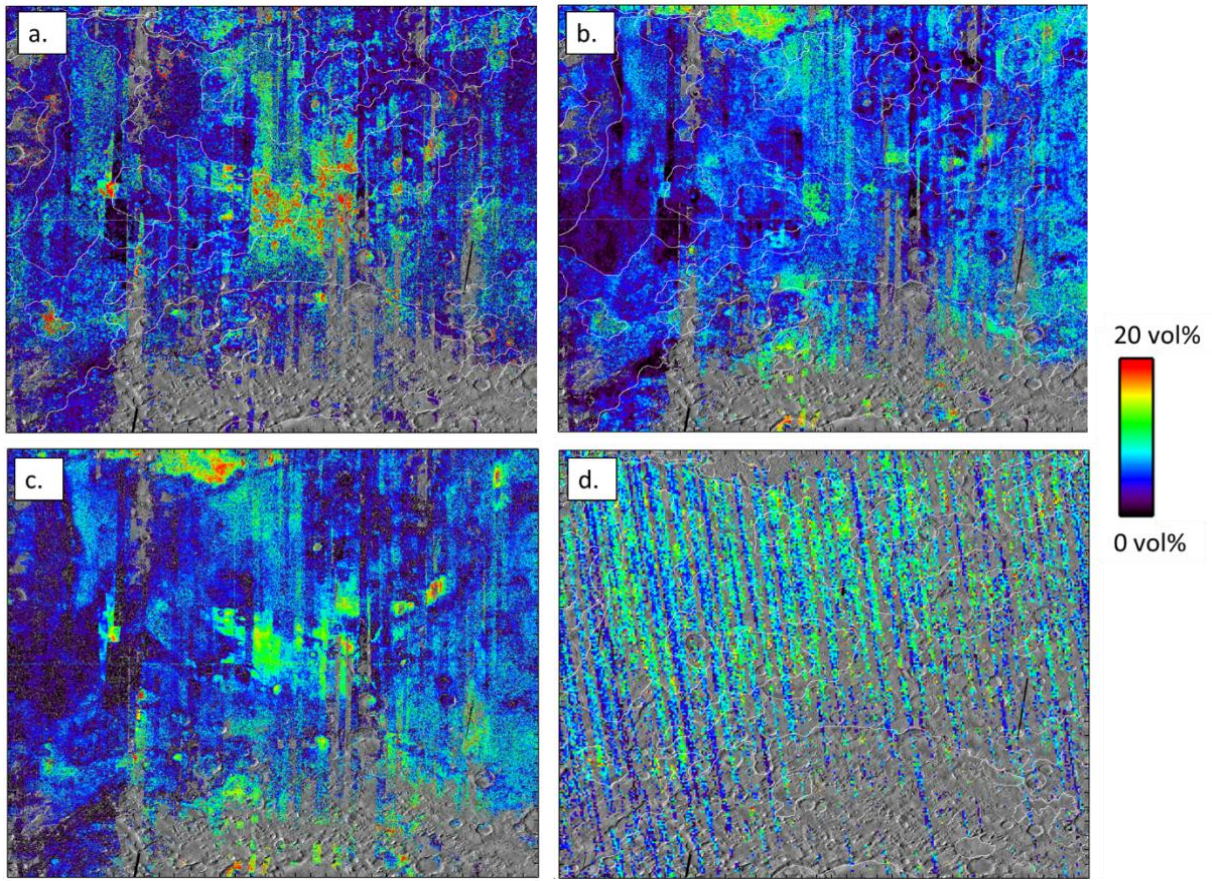

Figure 15 - Regional maps (45°S to 15°S in latitude / 300°E to 335°E in longitude) of basaltic glass and olivine abundances. (a) Basaltic glass when olivine is added. (b) Olivine when glass is included. (c) Olivine without glass. (d) TES olivine abundance map.

To confirm this observation and avoid the ambiguity, we rely here on the TES data for which TIR spectral signatures of olivine and glass are significantly different (Wyatt et al., 2001). This instrument indeed confirms that olivine has been detected in this region as well as in many others where the model retrieves a significant amount of olivine (Hoefen et al., 2003; Hamilton et al., 2003; Koeppen & Hamilton, 2008; McSween et al., 2006, Figure 15d). This example also illustrates the difference of spatial coverage and sampling between the two instruments, which can complicate a detailed comparison. We nevertheless consider that the OMEGA and TES identification of olivine at the surface of Mars justifies our hypothesis not to include basaltic glass in our numerical simulations. In the following, we thus chose not to present the basaltic glass component in the final global maps.

### 3- Global modal mineralogy

#### 3.1. Global maps

The global mineral abundances maps presented on Figure 16 represent final products of the Shkuratov-based modelling of the NIR OMEGA spectra over the Martian surface from -60° to 30°. These final products will be distributed to the community thorough the PSUP website in the near-future (http://psup.ias.u-psud.fr, Poulet et al., 2018). Uncertainties due to the fitting methodology, instrumental calibration and grain size can produce significant errors. However, we consider that both dataset and methods have been eventually demonstrated to be consistent and robust, allowing for producing global maps of individual phases. If we combine all the uncertainties discussed in the previous section, we find uncertainties on the abundances ranged between 4 and 14 vol%. The global uncertainty is calculating performing the square-root of the residual sum of squares of all individual uncertainty discussed in the previous sections (section 2.2 (Fig. 7), section 2.3 (Fig. 9), section 2.4 (Fig. 10), section 2.5, section 2.6 (Fig. 13), and section 2.7). Table 2 summarizes the averaged abundances derived from these global maps. On average, the surface composition is made of ~49 vol% of plagioclase,34 vol% of pyroxenes, and 4 vol% at most of olivine and and 14 vol% of dust. This mineralogy is similar to evolved volcanic (or gabbro) basalts confirming the previous regional investigations (Poulet et al. 2009b). In this table, we distinguish the errors extensively discussed in section 2, and the standard deviations that reflect compositional spatial diversity as it will be shown in the following sections.

The modelled pixels have been classified as a function of each major geological period (Noachian, Hesperian and Amazonian) by using the recent global geological map of Tanaka et al. (2014) adapted to the spatial resolution of the OMEGA data set. 82.5% of pixels are attributed to the Noachian period, 17.4% of the pixels to the Hesperian and less than 0.1% to Amazonian. The Tanaka map indicates that the most of the Amazonian terrains of our sample corresponds to pixels located in Valles Marineris terrains. These terrains appear to be dominated by dunes, but we have to keep in mind in the following that the origin of these dunes might not be representative of this geological period. Detailed analyses at local/regional scale would be necessary to better correlate geological units with modal mineralogy but such a work is out of scope of this paper, as the main objective here is to focus on global trend and planetary scale.

To start with, the spatial distribution of each mineral is reviewed in the following subsections.

|  | Dust | HCP | LCP | Plagioclase | Olivine | Magnetite |
|---|---|---|---|---|---|---|
| Average | 14 | 21 | 13 | 49 | 4 | 0.6 |
| Median | 3.81 | 24.22 | 14.40 | 49.72 | 3.83 | 1.33 |
| Standard deviation | 3.33 | 4.40 | 4.70 | 6.52 | 4.15 | 1.02 |
| Error | ± 12.5 | ±6.3 | ±6.2 | ±14.2 | ±4 | ±0.5 |
| Average grain size (µm) | 5 | 85 | 94 | 180 | 100 |  |
| Standard deviation (µm) | 0 | 55 | 69 | 54 | 0 |  |
| Density (g/cm³) | 0.8 | 3.3 | 3.3 | 2.7 | 3.4 | 5.2 |

Table 2 – Summary table of the results on abundances (in vol%) derived from all pixels mapped in Figure 16 and grain size ranges (µm) resulting from 8.7 million OMEGA spectra. The uncertainties are computed from the square-root of the quadratic sum of all individual uncertainties detailed in previous sections. The densities will be used to convert abundances from vol% to wt% in following sections. The density for the dust is the bulk density and the other densities are mineral densities. Mean, median, and standard deviation values over the global maps are listed for the 6 end-members.

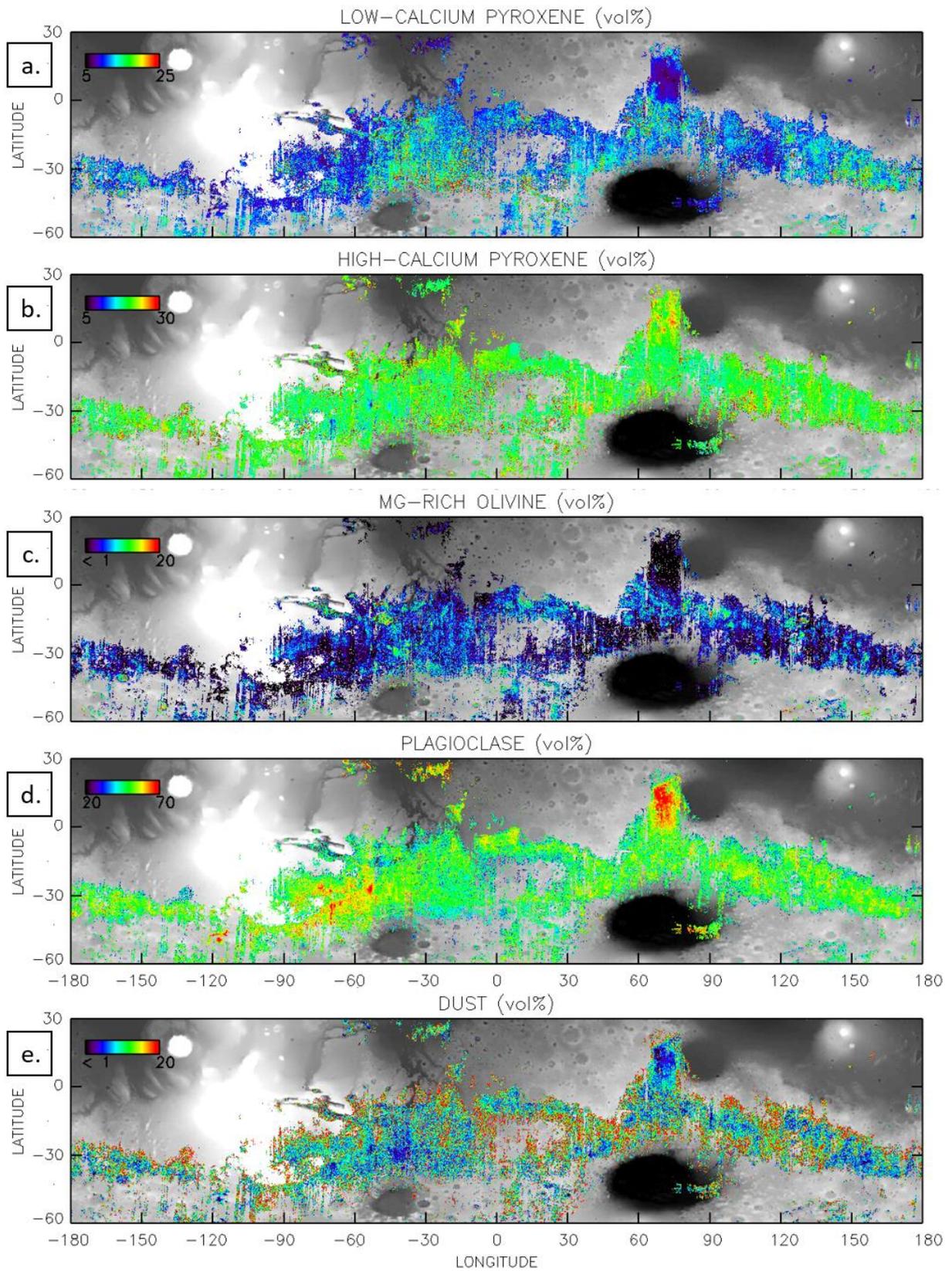

Figure 16 – Spatial distribution of abundance between 60°S to 30°N of (a) LCP. (b) HCP. (c) Olivine. (d) Plagioclase. (e) Dust. The abundances are in vol% and plotted on MOLA background.

In the following, we will compare the different mineral abundances with previous mappings of minerals (HCP, LCP and olivine) based on spectral indexes, and look for potential correlation between grain size and thermal inertia (section 3.5). The comparison with previous TES-based global abundances studies (Rogers & Hamilton, 2015, Koeppen & Hamilton, 2008) will be also presented (section 3.6). We will then apply standard classification method to modelled mineral distributions, which will allow identifications of mineralogical assemblages not necessarily recognized as distinct in previous studies (section 4). After calculation of chemical composition in section 5, we will examine the relationships with the GRS-based global distributions of some elements.

3.2. Plagioclase with magnetite inclusion

Plagioclase is the most abundant mineral that covers the low albedo equatorial terrains of Mars. It is widespread on the surface of Mars with more than 20 vol% at any location (Figure 16). The plagioclase distribution is relatively homogeneous on the surface with wide areas where the abundances reaches more than 50 vol%. However, we find regions where the total pyroxene (LCP+HCP) abundance dominates the plagioclase, thus this phase is not prominent on the entire surface. Globally, the spatial distribution of plagioclase seems to be anti-correlated with the olivine distribution, which could suggest at least two major mineral assemblages: a more evolved basaltic composition versus a more ultramafic composition.

Syrtis Major (early Hesperian volcanic unit) and Acidalia Planitia (late Hesperian transition unit) are enriched in plagioclase (up to 70 vol%) compared to the southern highlands. Most of the Thaumasia Planum also presents more plagioclase than the rest of the Noachian aged terrains that present on average ~49 ± 8 vol% of plagioclase. As an exception, a large portion of Bosporos Planum and Thaumasia Planum Noachian units present a large plagioclase abundance (~ 60 vol%) indicating that this end-member abundance may no be correlated with the surface age.

The grain size distribution for plagioclase is centered on 150 µm which corresponds to the initial value used in the model. However a wide range of grain sizes is explored from 5 to 450 µm for this end-member. No global correlation with the plagioclase abundance is highlighted regarding the grain size, we find relatively smaller grain sizes in Syrtis Major (120 µm on average) compared to Solis Planum (220 µm on average) which are both strongly enriched in plagioclase.

We recall that magnetite is included in the plagioclase end-member in the form of small inclusions (<10 vol%) to account for low albedo. The modelled magnetite abundance is anti-correlated with the albedo with abundance < 1 vol% for most part of the modelled spectra with albedo > 0.15 (Figure 17).

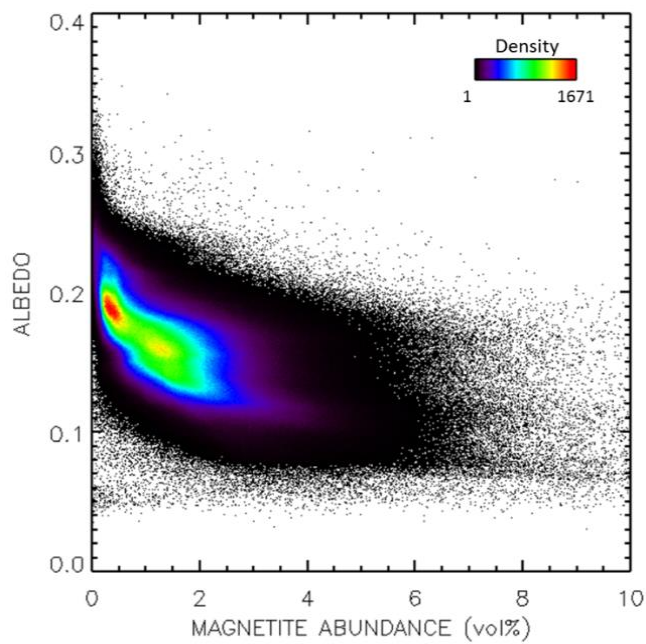

Figure 17 - Density plot of the abundance of magnetite with respect to the albedo.

### 3.3. Pyroxenes

#### 3.3.1. High-calcium pyroxene

HCP is widely spread on the overall surface with abundance values between 20 vol % and 35 vol % as shown on Figure 16 (b). The spatial distribution shows strong abundance variations at local scale inside wide well-recognized geological units. The highest abundances are found in parts of Hesperia Planum. Syrtis Major also appears to be locally enriched in HCP as shown by the zoom of this region (Figure 18). HCP is homogeneously distributed with abundance of ~21 vol% but local HCP-rich deposits of a few tens of pixels wide are observed. These local enrichments (up to ~35 vol%) correspond to crater ejectas dated from the Hesperian period (Baratoux et al., 2011). Meridiani Planum and Xanthe Terra also present high (> 30 vol %) HCP abundance. For the Noachian terrains the average value is about $20 \pm 4$ vol% and for the Hesperian terrains it is centered on $23 \pm 4$ vol%. The proportion of HCP seems to be increasing from Noachian to Hesperian except for South Acidalia that consitutes one of the youngest terrains in our sample but is depleted in HCP. Actually this terrain is quite unique because it is composed primarily of plagioclase and dust with poor amount of pyroxenes and olivine.

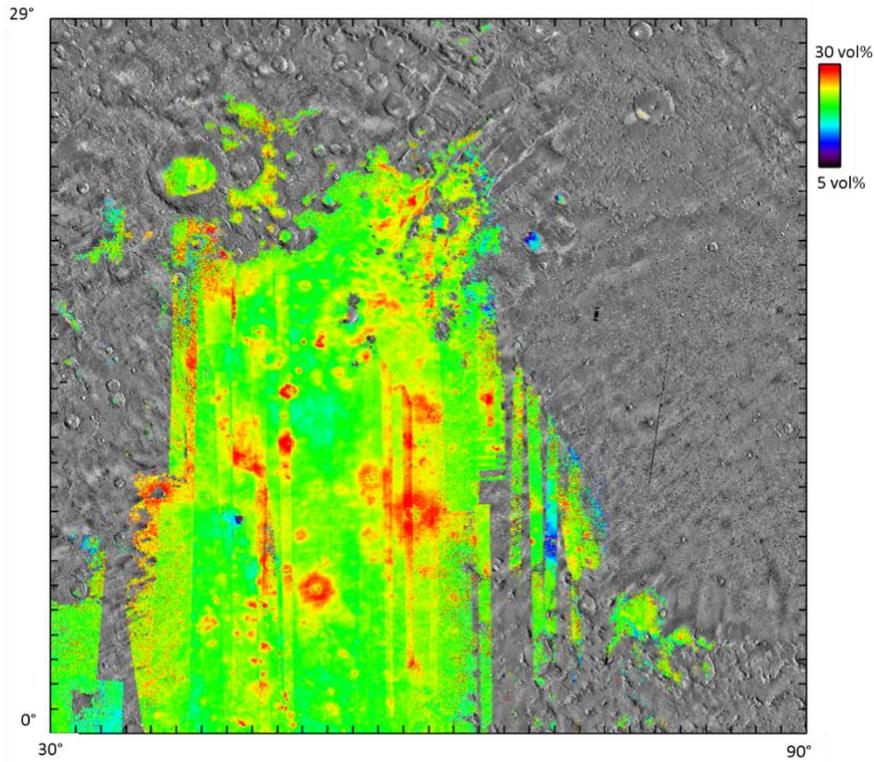

Figure 18 - Regional HCP abundance maps in Syrtis Major.

Finally, we notice that large values are found at the boundaries of the studied low albedo regions (Figure 16b, Figure 18). We cannot exclude that it could be the result of modelling artifacts. Actually, these regions exhibit small grain sizes, with values between 5 and 50 µm. They are also characterized by small band depth (Riu et al., 2019, companion paper 1). Yet the weight of the spectral characteristics of HCP is decreasing with decreasing band depth, making the fitting procedure less reliable. During the minimization procedure, the model attempts to decrease the grain size to fit the weak signature and the larger albedo. As a compensation, the abundance increases up to large values. Even though the RMS values fall under the acceptance threshold, those terrains have thus to be treated with caution. Overall, the grain size distribution for HCP is centered on 70 µm with values in the range between 5 and 450 µm.

3.3.2. Low-calcium pyroxene

LCP is widespread on the overall modelled surface with abundance always larger than zero, but with noteable variations. The average abundance is of 13 vol%, and it can reach up to values > 25 vol% in the mid-southern latitudes (30°S to 45°S). Noachis Terra is well visible with the highest abundance values (Figure 16a), while LCP appears to be less abundant in Thaumasia Planum and Aonia Terra. The LCP abundance in Noachis Terra region is not homogeneous. Although those local spatial variations are less well delimited than the HCP local variations, we are able to distinguish LCP-rich or LCP-poor deposits of a few tens of pixels (100-200 km²) (Figure 19). In this region, more widespread LCP-rich detections are also mapped near the pyroxene detection boundary close to -40° in latitude (Fig. 19). In the northern hemisphere, this mineral phase is on average less abundant with the lowest abundance found in Syrtis Major and Acidalia Planitia (<10 vol %).

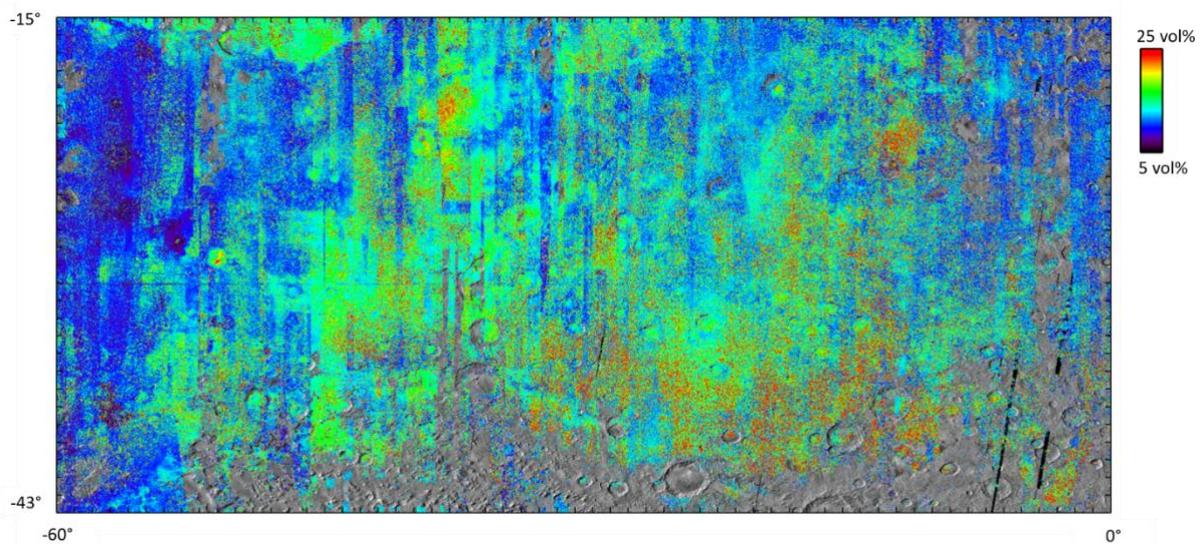

Figure 19 - LCP abundance map over the Noachis Terra region delimited by the quadrangle of coordinates: 45°S-15°S and 300°-360°E. Significant variations are observed with enrichment and depletion of LCP over zones of a few tens of OMEGA square pixels.

Although further work shall be done to better understand the geological context of these enrichements and depletions, the observed variations seems to be correlated with the age of terrains. From the geological map of Tanaka et al. (2014), we find the Noachian-aged terrains have an averaged abundance of 14 ± 4 vol% with detections up to the 25 vol% observed in Noachis Terra. The Hesperian-aged terrains exhibit LCP abundance of 11 ± 4 vol%. The variation of the LCP abundance with age is thus quite significative with a decrease in LCP abundance from older to younger terrains. Another view of this evolution is to plot the evolution LCP on the total pyroxene content (LCP/(LCP+HCP) ratio) with age as shown in Figure 20: a decrease in LCP is globally highligthed.

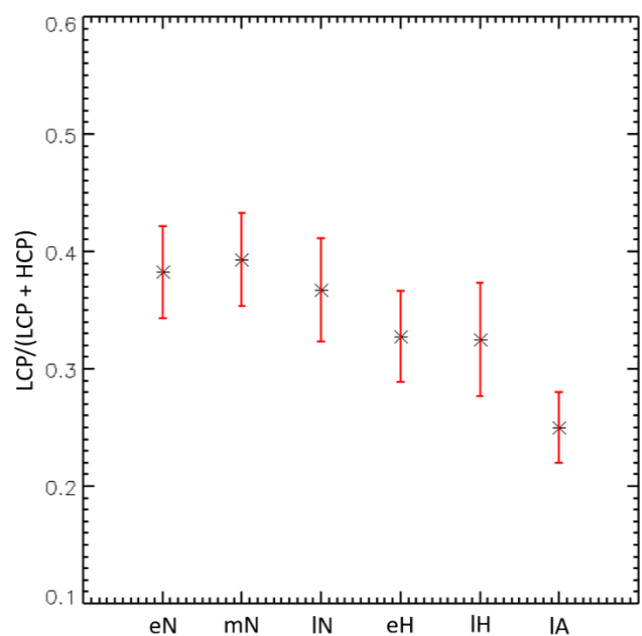

Figure 20 - Evolution of the LCP/(LCP+HCP) ratio with time (N = Noachian, H = Hesperian and A = Amazonian, e = early, m = middle, l = late. Terrains of the middle Hesperian, early Amazonian and middle Amazonian eras are not plotted because of the lack of points. The errors bars correspond to the standard deviation of the ratio for each considered epoch. At the global scale, the LCP/(LCP+HCP) ratio decreases with time.

The grain size distribution is centered around 80 µm which is smaller than the 100 µm that is set as an intial condition in the modelling. The range of values explored are between 5 to 450 µm approximatively. We observe a strong dependency between the grain size and the abundance: when the modelled abundance is large (> 20 vol%) the resulting grain size is always < 50 µm, for all regions discussed before. As for the HCP, these observations show that the model tends to decrease the grain size to fit weak pyroxene signature and larger albedo. The terrains presenting large LCP abundances will thus have to be treated with caution. The comparison with other orbital studies can be useful to decorrelate these effects (section 3.6).

### 3.3.3. Comparison with spectral index map

The derived abundances and grain sizes can be compared with the global distribution of pyroxene obtained thanks to a spectral index (Figure 1). The goal is to quantify whether the pyroxene index is correlated with the resulting abundances and to evaluate the impact of grain size parameter on the pyroxene index. To do so we sum the abundances of HCP and LCP and compare the value to the pyroxene spectral index that enables the detection of both calcium-poor and calcium-rich poles. For the comparison with the grain size both grain sizes are averaged.

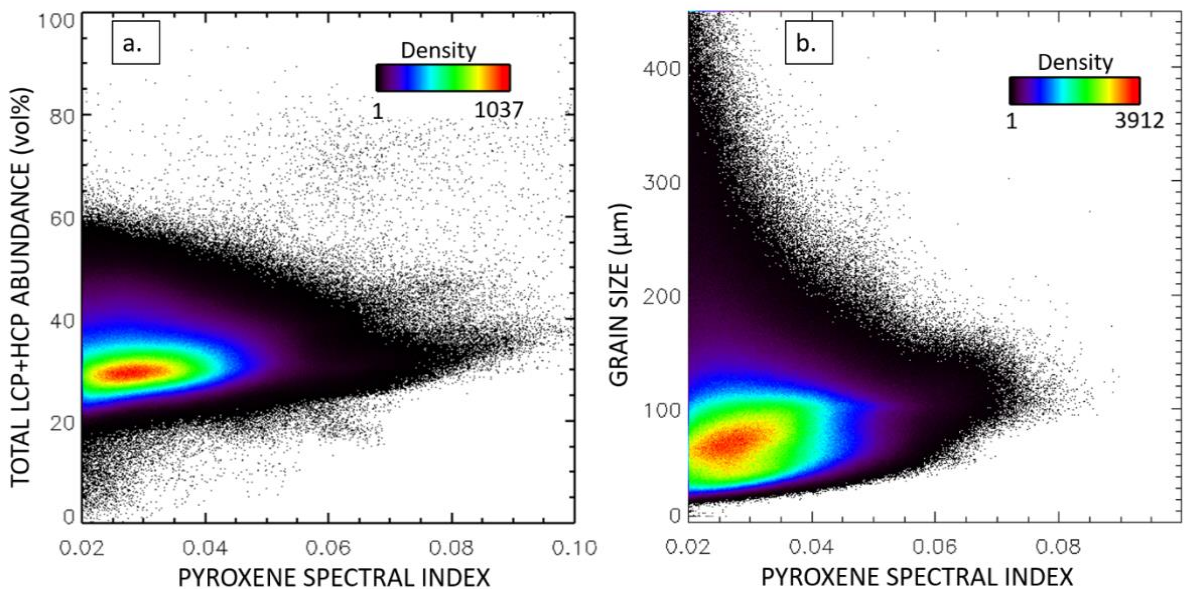

Figure 21 – (a) Density plot of total pyroxene abundance (LCP + HCP) with respect to pyroxene spectral criterion. (b) Same as (a) but for the average pyroxene grain size.

The total pyroxene abundance is almost always > 20 vol%, which tells us that threshold used for the spectral index at 2% enables the detection of pyroxene with abundance > 20 vol%. Surprisingly, the spectral index is not correlated with the pyroxene abundance (Figure 21a). This suggests that the spectral index is a good tracer for the detection of these minerals but it does not seem to be well representative of their abundances.

The grain size is expected to be correlated with the pyroxene band depth (i.e. the spectral index) with decreasing spectral index for smaller grains. It is shown on Figure 21 (b) that there is a positive correlation between band depth and grain size for sizes up to ~150 µm, which constitutes more than 99% of the data points. Nonetheless, for larger grains (> 200 µm) sizes the spectral index start to decrease (Harloff and Arnold, 2001). It thus appears that the grain size plays a more important role in the evolution of the spectral index in comparison to the abundance.

### 3.4. Olivine

We recall that (1) only pyroxene-bearing terrains (Figure 1) were selected for the modelling, excluding olivine-rich terrains such as Nili Fossae or the surrounding of Argyre basin (Poulet et al., 2009b; Ody, et al., 2013), and (2) the olivine end-member is a forsterite one with grain size fixed at 100 µm (section 2.1) (3) the modelled abundances depend on the spatial resolution of the observations, hence on the spatial sampling (here ~1.85 km/px). The olivine spatial distribution (Figure 16c) is significantly different from the other minerals distributions as large parts of the surface do not require this mineral. Actually, half of the modelled spectra exhibit olivine abundance < 3 vol%. Among these olivine-poor regions, there are Solis and Icaria Planum, as well as Syrtis Major and Xanthe Terra. Most of the detections > 5 vol% occur in the southern hemisphere. Abundance can reach up to 20 vol% with deposits scattered in Valles Marineris, North of Hellas basin, North of Tyrrhena Terra and South of Terra Cimmeria.

Noachis Terra presents a more homogeneous distribution with 8 vol% abundance on average that seems to be correlated with the LCP distribution.

The bottom terrains and the vicinity of Valles Marineris represent the largest olivine-bearing terrains, of our sample, with abundance close to 20 vol%. If we exclude this Amazonian region, the Noachian terrains overall cumulate three times more olivine detections of abundance > 15 vol% than the younger terrains, nevertheless the Amazonian units in Valles Marineris exhibit distributions that are clearly shifted to higher abundance values.

To test the sensibility of the model to this particular end-member and to determine a detection threshold, the model was launched on the overall surface with no olivine included in the mixture. We show that for the majority of the pixels, the RMS is not impacted when olivine is not included as an end-member in the modelling (Figure 22a). Those pixels correspond to abundance of olivine always < 3 wt% (Figure 22b) and the RMS for the simulation without olivine increases when the olivine is modelled with significant abundance (> 5 wt%) for the simulation with olivine. These observations show that (1) below 3 wt% olivine is not required and (2) for spectra where larger olivine abundance is modelled (> 3 wt%) the RMS is degraded when olivine is not included, which tends to confirm the presence of olivine in those terrains.

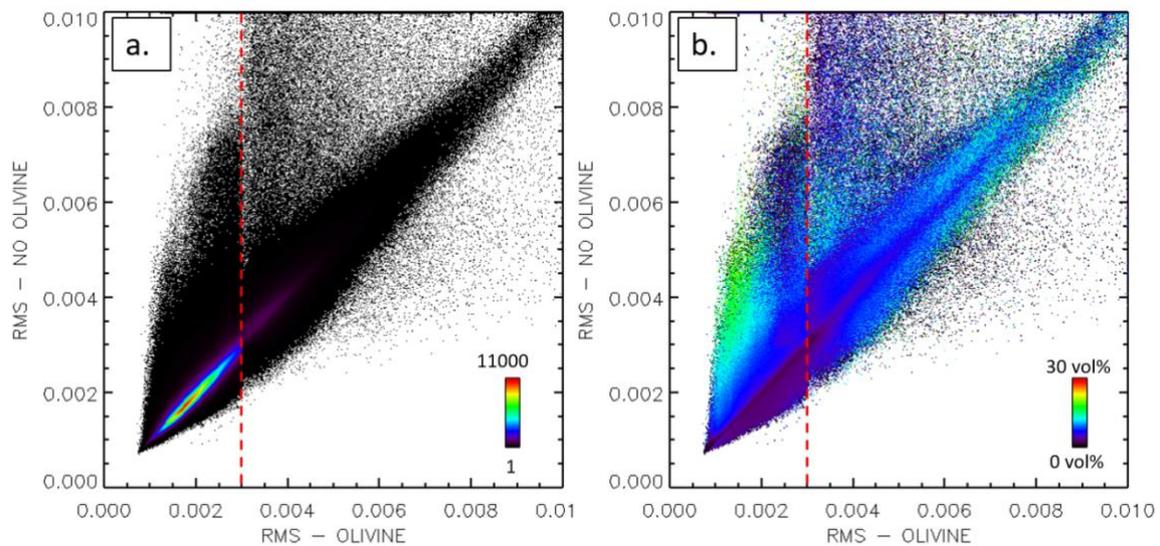

Figure 22 - RMS for the simulation with olivine versus the RMS for the simulation without olivine. The RMS values cover the entire surface. (a) Density plot. (b) Average olivine abundance for the simulation with olivine. The vertical red dotted line indicates the RMS threshold of 0.003.

As shown on Figure 10 (b) of companion paper 1, the olivine detection map shows only very localized olivine detections, which are in good agreement with the localized spots of olivine present in significant amount (typically > 10 vol%). We compare the abundance derived here for each pixel with the olivine spectral parameter (OSP1, Ody et al. 2013). A positive correlation between the spectral index and the abundance is highlighted (Figure 23). We can also see the threshold of the olivine index (set at 1.04) enables the detection of olivine grain of 100 µm with abundance ~1 vol%.

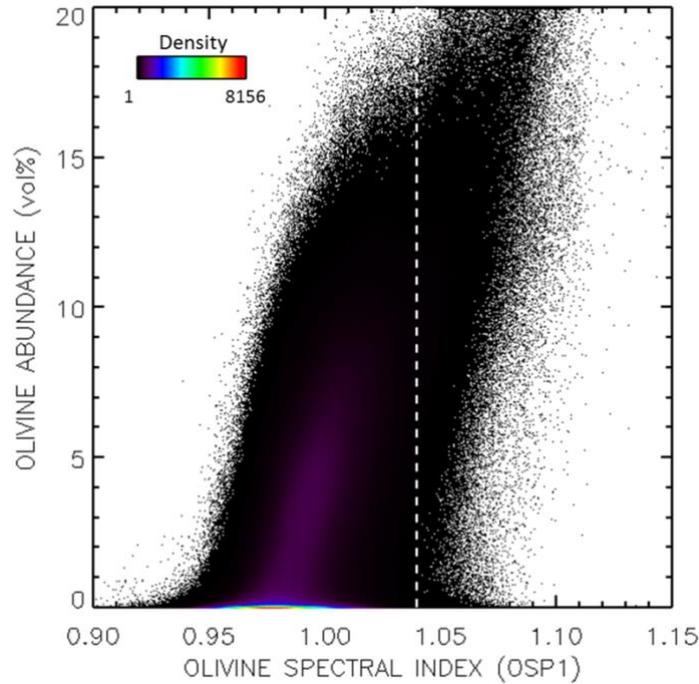

Figure 23 - Density plot of olivine abundance versus olivine spectral index referred to as OSP1 (Ody et al., 2013). The dotted line represents the 1.04 threshold detection limit of the OSP1 index.

Figure 23 also indicates that 75% of the data points for which the abundance is larger than 10 vol% are below the detection threshold limit (1.04). This demonstrates the advantage of the modelling over the mapping technique of spectral indexes. The modelling may reveal the presence of a significant amount of olivine in terrains where this mineral is not mapped using spectral indexes. This observation is also true for the plagioclase end-member that is modelled with abundance > 50 vol% for most of the surface without being mapped beforehand with spectral indexes. We show here that the model is capable of retrieving abundance of minerals when they are not detected based on their absorption bands and that it is thus necessary to use not only spectral indexes but also spectral modelling to access the surface composition.

3.5. Comparison with thermal inertia

The thermal inertia can provide insights into the physical properties of the surface such as average particle size. The goal here is to compare the OMEGA-based thermal inertia mapped at a global scale with a resolution of 4px/° (Audouard et al., 2014) with the derived grain size from the spectral modelling. The thermal inertia map covers the Martian surface from 45°S to 45°N, which restricts the comparison mostly to the southern hemisphere. For a regolith-like surface, the grain size is correlated to the thermal inertia: small (resp. large) grain sizes should be associated to small (resp. large) thermal inertia (Palluconi and Kieffer, 1981; Edgett and Christensen, 1991; Pelkey et al., 2001). Before comparison, we resampled the modelled grain size maps to a 4 px/° resolution. For each pixel, the comparison is performed with the average of grain sizes of the five end-members, weighted with the corresponding abundances. Figure 24 illustrates the dependence of this average grain size as a function of the thermal inertia.

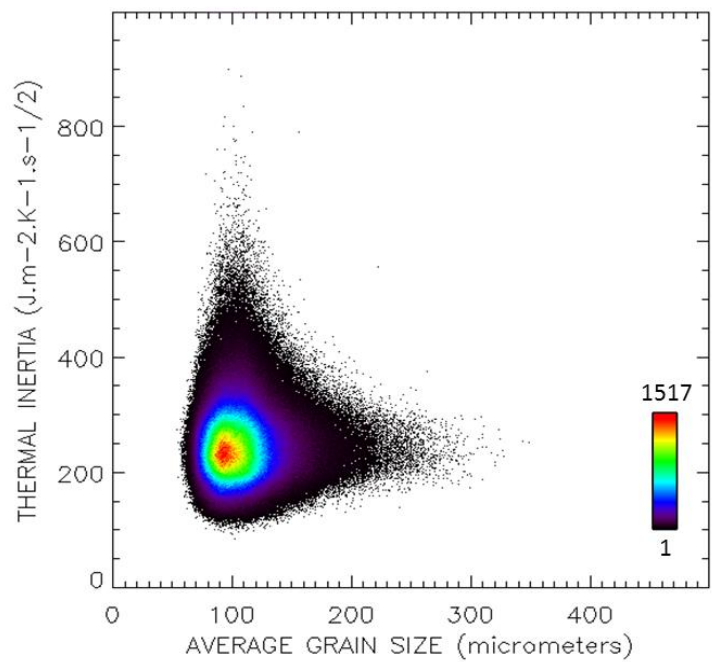

Figure 24 - Density plot representing the average grain size with respect to the thermal inertia for pixels located ranged between 45°S to 30°N. The derived grain size maps were resampled to 4 px/°, corresponding to the spatial resolution of the thermal inertia map.

Thermal inertia values for most of the studied low albedo regions range from ~150 to ~350 J/m/K/s$^{1/2}$, which corresponds to effective grain size from ~50 to ~1000 µm based on the thermal inertia/particle size relationship (Presley and Christensen, 1997; Pelkey et al., 2001). These thermal inertia values do not provide much constraints on the expected grain sizes. Although the grain size distribution centered on 100 µm is in the expected size interval, there is no positive correlation of the grain size with the thermal inertia. The lack of correlation may be due to the grain size averaging process and/or the degradation of the spatial resolution (from 32 to 4 px/°) required for the comparison.

Alternatively, previous detailed studies of the thermal behavior of the surface of Mars revealed the complexity of thermal behavior of the Martian surface with key roles of the dust and of surface (vertical, horizontal) heterogeneities at various scales (Audouard et al., 2014).

3.6. Comparison with TES-based mineral abundances

A global data set of surface mineralogy was determined by the TES instrument. It differs from the NIR spectral region in that (1) it is not primarily sensitive to Fe-bearing minerals (2) the materials to which this wavelength region is sensitive have strong absorption coefficients (3) there is no model-dependence as mineral abundances can be derived from linear deconvolution. On the other hand, the TIR is strongly sensitive to the atmospheric dust, which under high opacity conditions may dominate the useful surface signal. The TES-OMEGA comparison is performed with the most complete and best described TES-derived mineral maps from Koeppen & Hamilton (2008) and Rogers & Hamilton (2015). After private communication with D. Rogers, the TES maps selected for the spatial comparison are not the ones presented in Koeppen & Hamilton, (2008) but those used (but not shown) in Rogers & Hamilton (2015). Rogers and Hamilton (2015) used the mineral maps from Koeppen and Hamilton (2008), but, for each pixel, they normalized the mineral abundances to sum to 100% after excluding modelled atmospheric and blackbody contributions. As described by Rogers and Hamilton (2015), this normalization removes the dependance of overall spectral contrast, which in turn is dependent on grain size. Use of normalized values most closely represents modal mineral abundance, and is appropriate when conducting pixel-to-pixel (spatial) comparisons of modal mineralogy. These corrections significantly modified the spatial distribution. Due to its higher spatial resolution, the OMEGA abundance maps have first to be adapted to the TES resolution (8px/°). In addition, a mask was applied to both datasets to compare the exact same sample of pixels. Note also that the abundances can only be compared in terms of relative abundances because the set of end-members used in the deconvolution process of TES dataset is much larger. The TES abundances derived from maps of Rogers & Hamilton (2015) are thus renormalized to 100 vol% by keeping only the four end-members: plagioclase, HCP, LCP and olivine. While performing this normalization, we exclude the TES high-silica phase end-member that is a relatively abundant end-member in the final TES map (from 10 to 40%). This may result in higher uncertainties in the final relative TES abundances. The OMEGA binned maps are also renormalized excluding the dust component not taken into account with TES. This processing enables a more rigorous comparison in terms of spatial distribution and a relative comparison in terms of overall derived abundances. The corresponding normalized abundances are compared in detail on Figure 25. Globally, the TES derived abundances in our sample show the same type of mineral assemblages with plagioclase and pyroxenes as major components with relative average abundances of $45 \pm 7$ vol% and $46 \pm 6$ vol% respectively. The average abundance of olivine derived with TES ($9 \pm 4$ vol%) is larger than that extracted from the OMEGA dataset ($4 \pm 3$ vol%). But olivine is the minor phase in both cases. The main difference between the two instruments datasets is the LCP/(HCP+LCP) proportion. LCP is the dominant pyroxene phase with $32 \pm 9$ vol% on average compared to HCP with $14 \pm 7$ vol% for TES while we find average abundances of $15 \pm 4$ vol% and 24 ± 5 vol% respectively for LCP and HCP with the OMEGA rescaled maps. It is important to note here that reported TES abundances have evolved between the study of Bandfield (2002) and Koeppen and Hamilton (2008) due to the inclusion of new pigeonite and olivine end-members in the latter study (Koeppen and Hamilton (2008)), with LCP abundance evolving from 0 vol% to more than 30 vol%. However with the latest TES maps that we use here for the comparison (Rogers and Hamilton, 2015), if we combine the total pyroxene content (LCP+HCP) for both datasets, they look very similar (Figure 25c).

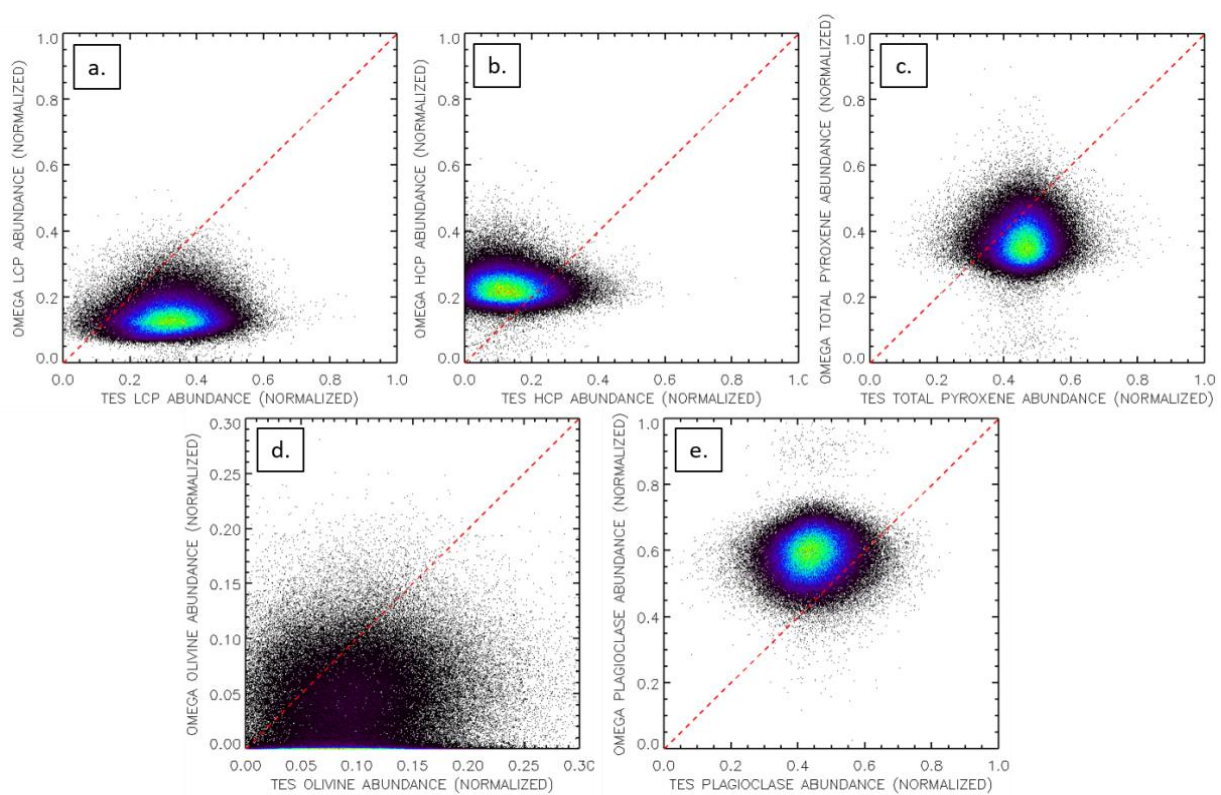

Figure 25 - Scattered plots of TES-derived abundances versus OMEGA-derived abundances in vol%. (a) LCP. (b) HCP. (c) Total LCP+HCP. (d) Olivine. (e) Plagioclase.

We now examine the correlations in terms of spatial correlation. Larger abundance contrasts are globally observed with OMEGA resolution, which could be due to the higher spatial resolution by a factor 5. This effect has to be kept in mind in the following, but we believe that general trends can still be discussed.

The comparison is of special interest for the plagioclase end-member because this mineral phase could a priori constitute a caveat for the modelling of OMEGA spectra due to the lack of strong absorption features in the NIR. On the contrary, it is detected with TES thanks to the presence of several fundamental vibrational modes in the TIR (Nash & Salisbury, 1991, Christensen, et al., 2000). Strong spatial correlations are found among the two datasets in Solis Planum and Noachis Terra. Solis Planum appears enriched in plagioclase in both cases where Noachis Terra presents rather lower plagioclase abundance. The main differences observed in the two sets of data are in Hesperian-aged terrains of Syrtis Major and Southern part of Acidalia Planitia (up to 30°N), which are dominated by plagioclase in the OMEGA maps. This observation is not in agreement with the TES feldspar abundance map that globally exhibits plagioclase content closer to the average in those regions.

For high-calcium pyroxene we find good correlation in Syrtis Major and North Acidalia that appear enriched and depleted in HCP respectively. In Hesperia Planum the OMEGA-derived abundance map shows some localized HCP-rich spots. Although this region appears also enriched in HCP with TES, it does not present the same contrast likely due to lower spatial resolution. Valles Marineris and its surroundings seem enriched in HCP in both cases.

The low-calcium pyroxene distribution is very similar: Syrtis Major and Solis Planum are depleted in LCP, Noachis Terra and Tyrrhena Terra present high LCP content and Meridiani Planum shows average LCP content. The good consistency for Noachis Terra reinforces the confidence on the OMEGA-derived LCP content as the abundance/grain size dependence (section 3.3.3) could constitute an issue regarding the large derived abundances in this region. The strongest difference is found inSouth,Acidalia where LCP is not present in the OMEGA map, while large content (37 vol% on average) was obtained from TES.

As for the LCP, the olivine distribution obtained with OMEGA is in very good agreement with the TES map. Olivine-bearing terrains are found in Tyrrhena Terra and Meridiani Planum. Enrichment of olivine is also present in Thaumasia Planum and south of Valles Marineris. North Acidalia, Terra Sirenum and Noachis Terra do not show significant olivine abundance. The good correlation in terms of spatial distribution for the olivine enables first to confirm the detections with high abundances (> 10 vol%) and low spectral index (< 1.04) (section 3.4) and second to exclude the possibility for the presence of basaltic glass (section 2.8). The major difference is found in Syrtis Major where we detect olivine in Nili Patera only, while TES-derived abundance in this region is quite homogeneous and close to the average olivine abundance (9 vol%) which illustrate the importance to go at higher spatial resolution which is the case with OMEGA. With a spatial sampling of 32 px/° (degraded to 8 px/° for the comparison), we are able to detect small olivine patches that are not highlighted with TES lower spatial sampling and thus we can better illustrate the spatial and compositional surface variability. This should also be the case for all other end-members modelled here.

## 4 – Classification of units by mineral abundances

### 4.1. Definition and geographic variability of mineralogical units

The resulting abundances maps were classified using a *k-means* clustering analysis to highlight different mineral assemblages and their spatial distribution (Tou and Gonzalez, 1974). The aim of the clustering is to classify the different abundances by minimizing the distance between the centroids of each class and their data points. This classification is based on five of the end-members, namely the abundances of plagioclase, LCP, HCP, olivine and palagonite. The distribution of magnetite and the grain sizes are thus not taken into account. The magnetite being imbedded in the plagioclase mainly to adjust the albedo, its spatial distribution is closely correlated to the albedo (Figure 17). Additionally, the grain sizes are less constrained than the abundances and adding them to the classification process would largely increase the number of parameters making the interpretations more complex without yielding statistical meaningful results. The *k-means* clustering aims at regrouping the different elements by maximizing the distance between the different classes for each parameter. The abundances with larger dispersion in terms of abundance values will thus have a bigger weight in the cluster formation. To prevent this bias, the abundances distributions were standardized beforehand so that their average is 0 and their standard deviation is 1. To do so, we subtracted the average value to each parameter and divided by the standard deviation reported in Table 2. The abundances were converted from vol% to wt%, using the densities in Table 2, to facilitate the comparison with *in situ* and meteorite-based measurements. The conversion into wt% depends on the mineral densities used (bulk or mineral density). Several ranges of densities were tested to evaluate the sensibility of this parameter (from 0.8 to 1.6 g/cc for the dust (Allen et al., 2001) and 3.4 g/cc to 1.6 g/cc for the other minerals). We found differences in the resulting abundances of the order of 1% at most for all mineral (section 4) and all oxides (section 5).

The resulting clustering consists of 7 classes as shown in Table 3 and Figure 26.

|  | Class 1 | Class 2 | Class 3 | Class 4 | Class 5 | Class 6 | Class 7 |
|---|---|---|---|---|---|---|---|
| % of pixels | 13 | 21 | 22 | 11 | 10 | 18 | 5 |
|  | 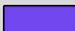 | 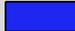 | 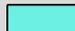 | 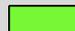 | 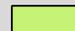 | 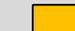 | 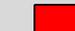 |
| Dust | 3.8 ± 2.3 | 3.3 ± 1.5 | 3.6 ± 1.5 | 2.9 ± 1.7 | 3.8 ± 1.9 | 8.1 ± 2.0 | 13.9 ± 3.2 |
| HCP | 22.6 ± 3.3 | 22.3 ± 2.6 | 25.0 ± 2.8 | 23.1 ± 3.5 | 33.0 ± 4.2 | 25.0 ± 2.6 | 26.2 ± 4.0 |
| LCP | 15.7 ± 2.8 | 13.8 ± 2.4 | 11.9 ± 2.5 | 24.9 ± 4.5 | 13.6 ± 3.5 | 15.2 ± 2.4 | 18.0 ± 3.0 |
| Olivine | 11.7 ± 3.0 | 6.6 ± 2.0 | 1.1 ± 1.4 | 3.6 ± 2.9 | 3.5 ± 3.1 | 1.9 ± 2.1 | 2.4 ± 2.9 |
| Plagioclase | 45.0 ± 3.7 | 52.2 ± 2.9 | 56.4 ± 3.2 | 43.6 ± 4.2 | 45.0 ± 4.6 | 48.9 ± 3.0 | 38.9 ± 5.4 |
| % of pixels |  |  |  |  |  |  |  |

| Noachian | 88.8 | 87 | 76.99 | 90 | 82.7 | 89.98 | 87.01 |
| Hesperian | 11 | 12.98 | 23 | 9.97 | 17 | 10 | 13.02 |
| Amazonian | 0.2 | 0.02 | 0.01 | 0.03 | 0.3 | 0.02 | 0.07 |

Table 3 – Average abundances of each class and standard deviation (±Δ) resulting from the k-means classification. The abundances are given in wt%. The percentage of pixels per geological era is also indicated.

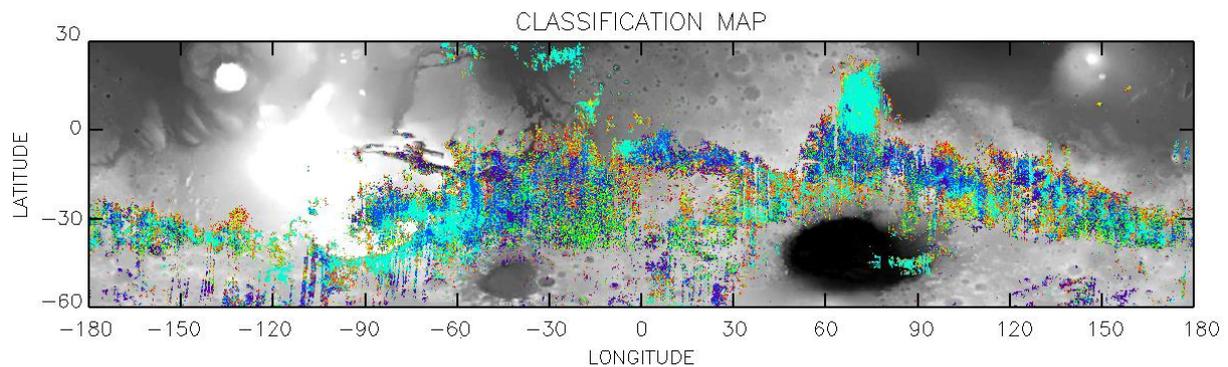

Figure 26 – Original classification map, before reduction of the sample, as described below.

The values of average abundances of some classes may be similar. This trend combined with the large standard deviations suggests that the abundances of each unit could be poorly constrained. As a consequence, we decided for each class to keep only the pixels were the abundances are found within ± 1 standard deviation from the average abundance. This filtering step reduces the sample to only 33% of the pixels but enables to better spatially differentiate the classes. This reduced sample will be used for all further analysis presented in the following sections.

Class 1 is a small class corresponding to terrains strongly enriched in olivine in comparison to the overall surface. The Valles Marineris region defines the region that is spatially the most distinct at global scale. Class 2 is found in part of Thaumasia Planum, Tyrrhena Terra and Sinus Meridiani (east portion). The average abundances of class 2 are well representative of the average abundances found on the entire surface. Class 3 is the dominant class with 22 % of the modelled pixels (Table 3). This class is characterized by its high plagioclase content and low olivine and LCP contents. It is found in Syrtis Major and South of Acidalia Planitia. A large portion of Noachian aged terrains of Protei Regio, Bosporos Planum and Aonium Terra are also attributed to this class. This class thus highlights numerous terrains of distinct ages and natures that share the same high plagioclase content, which questions their distinct origin. Class 4 is mainly found in Noachis Terra and Terra Cimmeria. Its high LCP content (~25 wt%) is consistent with the age of those terrains (Noachian). It is this only class where the LCP content is higher than HCP. On the contrary, class 5 is strongly enriched in HCP with average content > 30 wt%. This class only accounts for 10% of the modelled pixels kept after filtering and is spatially correlated to Hesperia Planum and Terra Meridiani. The last two classes (class 6 and 7) differ from the rest of the classes with their high dust content. Two classes were likely clustered due to the difference in terms of plagioclase. But the pixels attributed to those two classes are mainly found on the edges of the low albedo studied terrains, which is consistent with their high dust content.

The 7 classes can be differentiated by their distinct mineralogy but they do not significantly vary in terms of surface age except for class 3 in which more Hesperian terrains are found than in the other classes (Table 3). This is likely due to the fact that the sampled terrains modelled in this study are mostly Noachian. To correct this trend, we decide to normalize the number of pixels of each class per era, as represented on the pie charts of Figure 27. Class 1 (olivine-enriched unit) accounts for half of the pixels associated to Amazonian. As previously pointed out (section 3.4), those points are located in Valles Marineris. But the small percentage of Amazonian terrains (< 1%) and the poorly-constrained origin of these terrains do not permit to conclude whether this trend is representative of the overall Amazonian period or associated only to our sample. In summary, the entire surface is characterized with distinct mineralogical assemblages but no clear trend in terms of temporal evolution is highlighted, except for class 3.

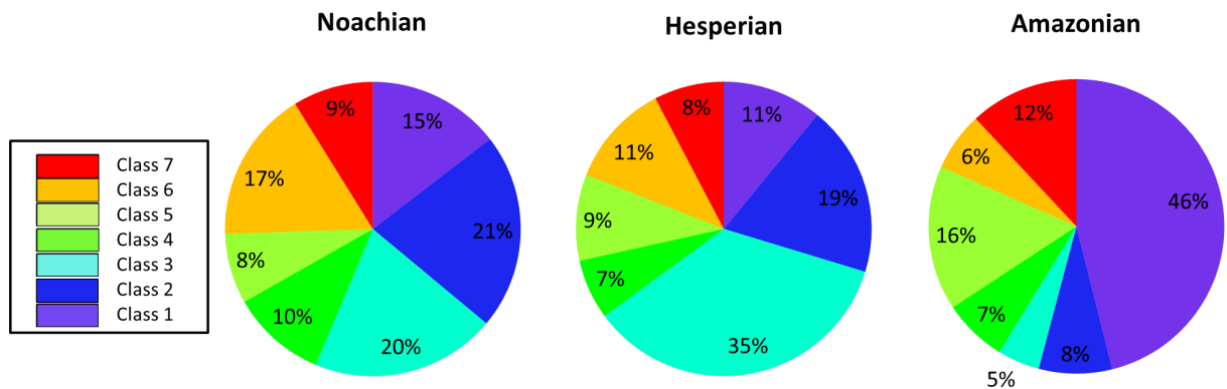

Figure 27 - Pie chart representing the percentage of pixels associated to each class for the three major geological eras. The number of pixels per era is normalized. The Amazonian pie chart is included, however our sample lacks of Amazonian terrains spread on the overall surface (only 0.1%, located mainly in Valles Marineris) which suggests a poor representativeness of this geological era.

4.2. Chemical composition of mineralogical units

As the spatial distribution of each class is defined, we can explore their variability in terms of chemical composition. Although the chemical composition from OMEGA modal mineralogy will be derived in the next section, we prefer to use here the GRS data, because they provide access to more elements such as K and Th (Boynton et al., 2007). Moreover, the GRS measurements probed in principle materials over several tens of cm in depth. The two data sets were adapted in order to enable the quantitative comparison. The OMEGA classification map (Fig. 26) was degraded to GRS spatial resolution ($5 \times 5°$ per pixel), while the GRS maps were masked to be consistent with the spatial coverage of OMEGA maps. Then the abundances of each element obtained for each class was normalized to the mean value of all pixels.

Except for classes 1 and 3, the classes do not significantly differ from one another and are all close to the average GRS abundances (Figure 28). Class 1, enriched in olivine, appears strongly depleted in Si compared to the others, which is consistent with its more mafic composition than the other classes. The poor Fe content could imply that the olivine is preferentially Mg-rich. This observation validates the choice of a Mg-rich olivine as an end-member for the modelling. Conversely, class 3 (enriched in plagioclase) appears enriched in Fe and Si and depleted in incompatible elements K and Th in comparison to the other classes (light blue on Fig. 28). This class is the only one exhibiting lower content in both K and Th than the average abundances. This particular observation could be related to the surface age of large portion of class 3 that regroups most Hesperian terrains (Table 3). The depletion in both Th and LCP for this particular class is nevertheless contradictory. Th is highly incompatible, which implies that its concentration should be inversely proportional to the degree of partial melting (Baratoux et al., 2011). Hence, a depletion in Th should be associated with an enrichment in LCP, as both supposedly trace an increase in the degree of partial melting. On the other hand, Baratoux et al. (2011) suggest that the concentration of incompatible elements in the mantle may be very sensitive to the previous melting, which could explain the depletion of Th observed in the younger terrains. Therefore, those mineralogical and geochemical trends may be the result of a combination of complex magmatic episodes that evolved with time.

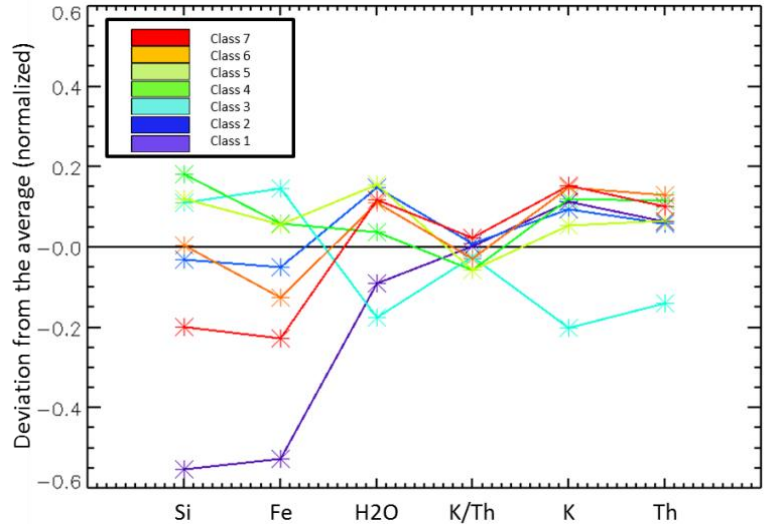

Figure 28 – Normalized variations of GRS abundances of 5 elements for each OMEGA-based mineralogical class with respect to the average value of each element on all pixels. The average and standard deviation for each element on all pixels are $20.2 \pm 2.5\%$ for Si, $13.3 \pm 2\%$ for Fe, $3.6 \pm 1\%$ for $H_2O$, $0.32 \pm 0.05\%$ for K and $6.10^{-5} \pm 1.10^{-5}\%$ for Th.

### 4.3. Comparison with previous orbital and *in situ* investigations

A certain heterogeneity in terms of mineral assemblages was highlighted thanks to the classification maps and the merging with GRS data in the previous sections. However, the fitting method that uses a narrow number of end-members and the relatively large OMEGA footprint (~1.5 km/px) could limit the constraints on the petrological processes. We thus propose below to compare the modal mineralogy obtained from the classification to smaller scale analyses (*in situ* and/or meteorites data) that sample the surface with higher spatial resolution.

Globally, the classification shows the same trend that the ones that were found by Poulet et al. (2009b) who investigated mineral abundances from modelling of selected local deposits (Figure 29, green triangles). Terra Tyrrhena (class 2) is found at the boundary between class 1 and class 2, which is consistent with its enrichment in olivine. Class 4 (enriched in LCP) is close to the small LCP-enriched terrains previously modelled by Poulet et al. (2009b) (labelled LCP on Fig. 29). Lavas in Syrtis Major (SML) and Hesperia Planum (HP), previously recognized as enriched in HCP, are found in class 5 that is also enriched in HCP. These good agreements between previous local studies and global mineralogical units demonstrate that the statistical classification method applied to the overall surface may be relevant to extract more local compositional information. However, there are some discrepancies especially with the Syrtis Major volcanic edifice. In this work, this structure exhibits one of the largest plagioclase abundances on Mars, while the previous modelling based on a smaller surface did not emphasize such enrichment in plagioclase. After the filtering process (excluding pixels for each class at ± 1σ), olivine-rich local spots such as Nili Patera does not belong to any mineral class and thus do not appear on Figure 29. Note that the local analyses performed for Nili Fossae terrains (referred to as SMN1 and SMN2) are also not a part of any class (SMN1 and SMN2, on Fig. 29). As the selection of the pixels used for the modelling was based on pyroxene-rich material excluding olivine-rich deposits such as Nili region for which the pyroxene spectral index is negative.

In comparison to the unaltered rocks investigated by *in situ* observations in Gale and Gusev craters, the mineralogy obtained with the OMEGA global modelling predicts less olivine. The olivine is a minor phase and its depletion relatively to the pyroxenes could be the indicator of a global depletion in olivine due to various processes such as global subsequent alteration episodes and/or formation of the Martian crust from a globally highly differentiated magma. In addition, as mentioned previously, the method of mapping and selection of pixels to be modelled excluded numerous olivine-bearing spots. Also the merging process during the creation of the global hyperspectral cube may have affected the olivine signature that is the most sensitive to the degradation of the spatial resolution and atmospheric load (Figure 5 of Riu et al., 2019). Note finally the olivine-bearing terrains have been extensively studied by Ody et al. (2013) who demonstrated the importance to select the best observations for identifying the signature of olivine and to careful map the spatial extension of its signature so as to constrain the formation of olivine-bearing magma.

Meteorites show a wide variety of compositions in terms of pyroxenes, olivine and plagioclase, which cover a large range of abundance values (Fig. 29). We find that only compositions of basaltic shergottites match OMEGA abundances as already emphasized by previous studies (Poulet et al. 2009b, Ody et al. 2015). Other SNCs are strongly enriched in LCP and olivine (picritic shergottites, chassignites and ALH84001) in comparison to the average composition of Martian terrains.

We here show the complementarity between local and/or *in situ* analyses and global analyses, as these multiscale studies can provide different mineralogical trends. The distinct mineralogical classes of OMEGA are in good agreement with local analyses where the local regions are globally spatially homogeneous. Conversely, the heterogeneity derived for small km-sized deposits and at microscopic scale is not observed at global scale with a 1.5 km size pixel. Such multiscale differences are not surprising as they have been already reported by other studies (McSween et al., 2009a; Sautter et al., 2015) However, it seems that our classification process tends to highlight global compositional trend at the expense of local variations. As already reported in section 3.3, we did detect some interesting local variations in Syrtis Major or Noachis Terra for example. But a more in-depth analysis of local compositional variations will have to be performed in order to extract local/regional trends and try to identify at what scale the heterogeneity that is encountered *in situ* can be perhaps highlighted.

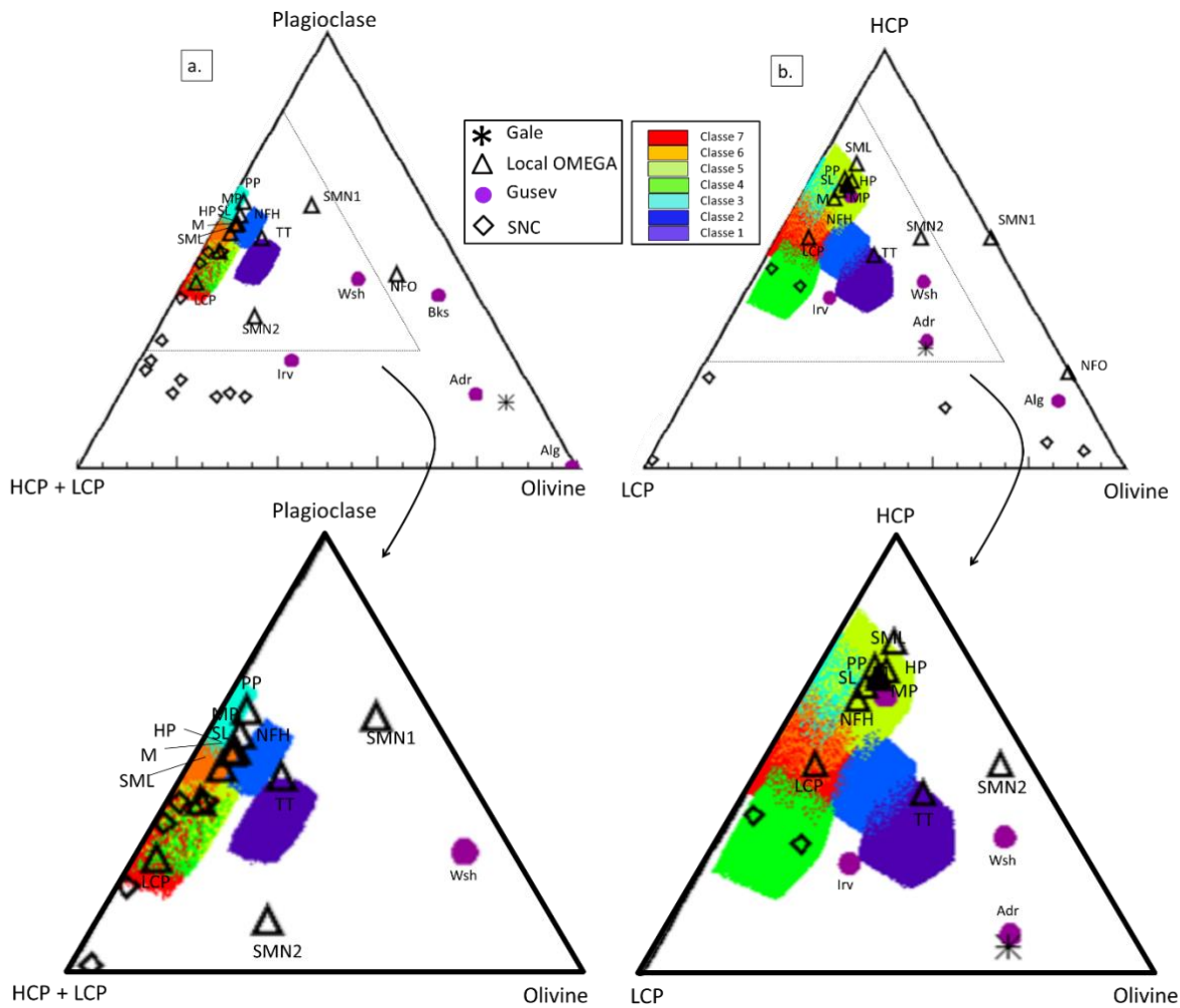

Figure 29 - Distribution of the abundances within the mineralogical classes projected in two ternary plots. (a) Plagioclase, LCP+HCP, olivine. (b) HCP, LCP, olivine. The abundances extracted from local OMEGA analyses (Poulet et al., (2009b)) are reported in the form of black triangles : M (Meridiani), HP (Hesperia Planum), MS (Mare Sirenum), TT (Terra Tyrrhena), PP (Protei Planum), SL (Solis Lacus), SML (Syrtis Major Lava), SMN1/2 (Nili Patera), NFH/NFO (Nili Fossae), LCP (LCP-rich terrains in the southern hemisphere). Gusev data (Adirondack (Adr), Irvine (Irv), Algonquin (Alg), Wishstone (Wsh), Backstay (Bks)) are represented by purple dots (McSween et al., 2009a, McSween et al., 2006). Gale crater point corresponds to the non-altered samples referred to as rocknest in Vaniman et al. (2014). It is represented by a black asterisk. SNC data are represented by black diamonds (McSween et al.,

1996, Hamilton et al., 2003, Warren et al., 2004). Only Shergottites are presented on (a). Bottom figures (c) and (d) are zooms of the top figures.

### 5- Oxide composition

This section presents the average oxide composition of the Martian surface calculated from the mineral abundances. In doing so, we are able to quantitatively compare the OMEGA data with chemistries measured *in situ* and in Martian meteorites. But we have to keep in mind that such chemistry is less diagnostic than modal mineralogy because assumptions about the partitioning of elements into minerals and the lack of secondary phases are required. The chemical composition is obtained by using the chemical composition of each end-member included in the mineral mixture. The conversion is reported in Table 4. Except for palagonite which chemistry comes from Allen et al., (1997), the chemical composition of other end-member comes from Poulet et al. (2009b). Wt % oxides for each phase are multiplied by the modelled weight fraction of that phase and combined to produce the derived oxide composition for each pixel.

|  | Palagonite (Dust) | Augite (HCP) | Pigeonite (LCP) | Labradorite (Plagioclase) | Forsterite (Olivine) | Magnetite |
|---|---|---|---|---|---|---|
| $SiO_2$ | 43.7 | 50.35 | 49.99 | 51.76 | 40.87 | 0 |
| $TiO_2$ | 3.8 | 0.35 | 0.53 | 0 | 0 | 0 |
| $Al_2O_3$ | 23.4 | 2.21 | 6.21 | 30.84 | 0 | 0 |
| FeO | 3.5 | 16.18 | 16.56 | 0 | 9.77 | 31.03 |
| $Fe_2O_3$ | 11.8 | 1.69 | 0.58 | 0 | 0 | 68.97 |
| MnO | 0.3 | 0.37 | 0.26 | 0 | 0 | 0 |
| MgO | 3.4 | 11.1 | 22.31 | 0 | 49.36 | 0 |
| CaO | 6.2 | 17.93 | 3.99 | 13.36 | 0 | 0 |
| $K_2O$ | 0.6 | 0.03 | 0 | 0.17 | 0 | 0 |
| Na$_2$O | 2.4 | 0.23 | 0.06 | 3.86 | 0 | 0 |
| Total | 99.01 | 100.4 | 100.49 | 100 | 99.99 | 100 |

Table 4 – Summary table of the chemical compositions of the different end-members used for the conversion from mineralogical composition to chemical composition.

On the igneous classification diagram that compares abundances of wt % silica (SiO$_2$) and wt % alkalis (Na$_2$O+K$_2$O), the OMEGA-based global chemistries plot in an oval, bullseye-type, density cloud, which overlaids basaltic rock class (Figure 30). Some pixels are also associated to more mafic rock (picritic type).

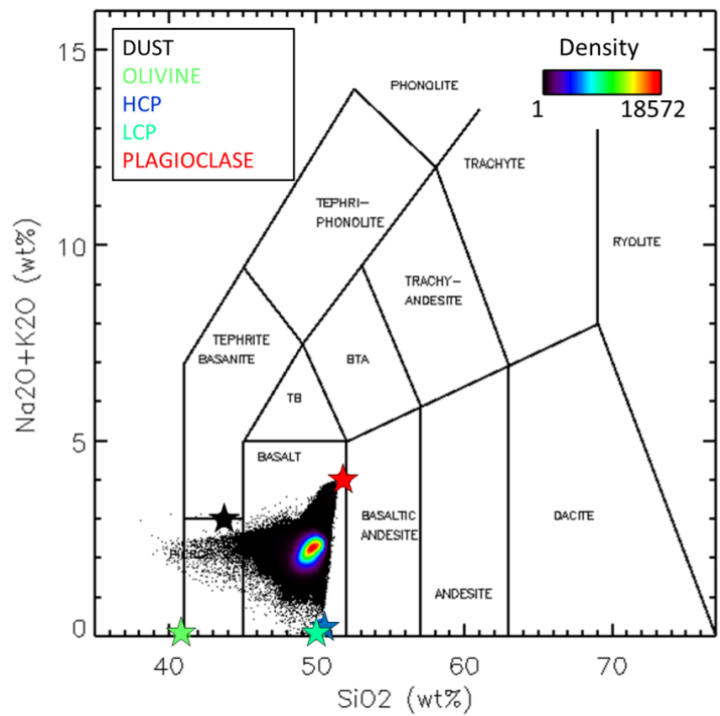

Figure 30 - Silica vs. alkalis density diagram from the modelling of the mineralogical abundances of pyroxene-bearing regions from OMEGA dataset. Data are overlaid over major rock classes. The colored stars represents the individual compositions of each end-member.

The OMEGA-based igneous classification is in good agreement with GRS measurements as well numerous *in situ* studies and Martian meteorites (McSween et al., 2009a,

Sautter et al., 2015). No basaltic andesitic composition is observed. The $SiO_2$ concentration cannot reach values higher than 51.76 wt%, because the chemical composition is directly related to the choice of end-members narrowing the range of possible concentrations of elements. This explains the sharp upper delimitation on Figure 30.

The abundance distributions of the minerals modelled (Figure 16) leads to a homogeneous surface in terms of Si composition with values between 21 and 24 wt%. Additionally, OMEGA-derived major element maps display geographic variations that correlate well with the variations in lithology observed previously (section 3). This expected trend combined to the uncertainties and large discrepancies in terms of spatial resolution makes difficult the spatial comparison to GRS global maps. In the following, we preferred to discuss the igneous chemical composition from a global point of view rather than compare the spatial distribution of both datasets.

Additional information can be revealed by the Al, Mg, and Ca contents. Ratioed abundances are superimposed to orbital, *in situ* and Martian meteorites measurements on Figure 31. Similar variations are observed: Ca/Si decreases with Mg/Si (Fig. 31b) and Mg/Si decreases with Al/Si (Fig. 31a). The Ca/Si vs Ms/Si slope is remarkably consistent with the regression line derived by McSween et al. (2009a) from various sources. OMEGA data are however plotted with an offset of about 0.05-0.10 in comparison to the line, which indicates an enrichment of Ca. Because Martian meteorites are depleted in Al relative to terrestrial rocks, Al depletion has been proposed as a geochemical discriminant for the whole Martian crust. However, such a trend has been put into question by further *in situ* and orbital observations (McSween et al., 2009a). The OMEGA global average data also support higher Al/Si ratio. The observed enrichment in Al and Ca predicted with the OMEGA data can be explained by the plagioclase and dust endmembers which are both enriched in Al and Ca (black and red stars on Fig. 31 (a) and (b)).

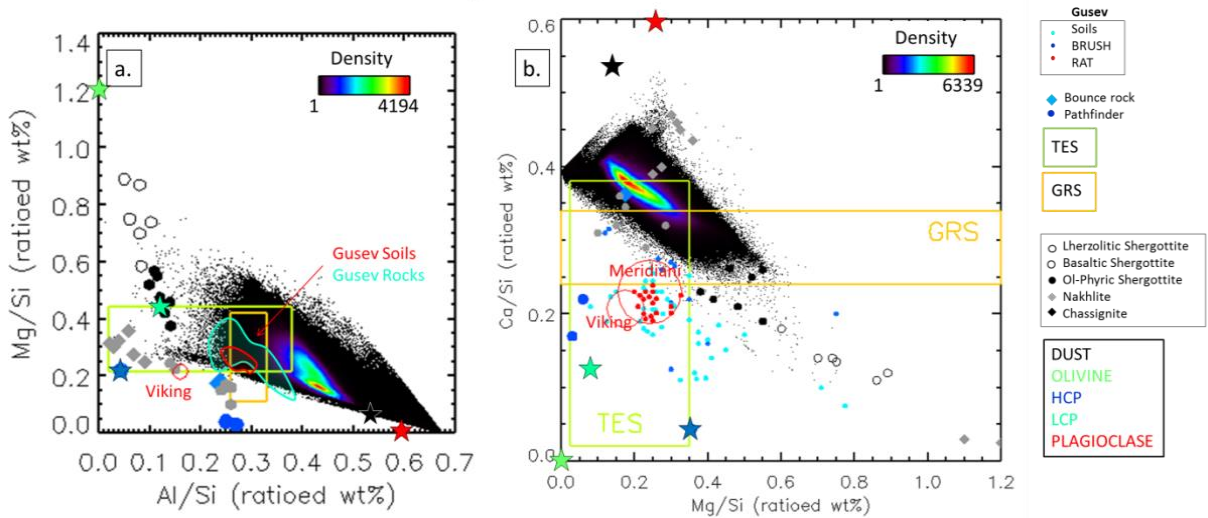

Figure 31 - (a) Mg/Si with respect to Al/Si, OMEGA data are represented in density. (b) Ca/Si with respect to Al/Si, OMEGA data in density. The GRS and TES data, *in situ* and SNC data are extracted from McSween et al. (2009a) and McSween et al. (2009b). The colored stars correspond to the individual end-members.

The high iron content of Martian meteorites was also commonly assumed to be fingerprints of Mars. Such a large value was later confirmed at a global scale by GRS value. This questions the low abundance of Fe (average abundance of 6.5 wt%) derived from the OMEGA-based lithology (Figure 32a). The difference between the two orbital data sets could be attributed to the fact that GRS samples the surface with depth of a few tens of cm in comparison to OMEGA which samples only the few first microns. Additionally, as for the modal composition, the oxide composition could be sensitive to the spatial sampling that is strongly improved with OMEGA (500 km/px for GRS and ~1.85 km/px for OMEGA). Further iron contents measured from meteorites and *in situ* in Gusev crater by the Spirit rover actually show a broader range of abundances than the one predicted by GRS (Figure 32a). The Fe content measured in Gusev is indeed ranged from 8 to 18 wt%. Although these values partly cover the OMEGA-based iron concentration, they are still significantly different.

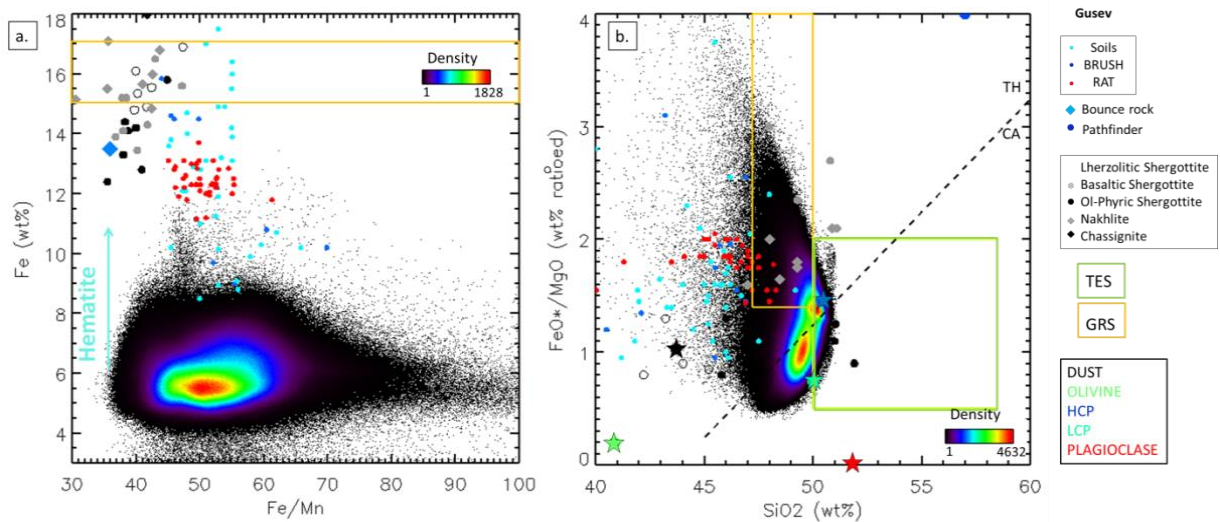

Figure 32 - (a) Fe with respect to Fe/Mn ratio. OMEGA data are represented in density. The vertical blue line indicates the trend when hematite is used instead of palagonite as the dust analog. (b) SiO2 with respect to FeO*/MgO ratio. The black dotted line shows the separation between dry tholeiitic (indicated as TH) and wet calc-alkali (indicated as CA). The GRS and TES data, *in situ* and SNC data are extracted from McSween et al. (2009a) and McSween et al. (2009b). The colored stars correspond to the individual end-members.

This discrepancy has been already discussed from the modelling of local selected regions (Poulet et al. 2009b). The iron content is very dependent on the total iron oxide content present in plagioclase, which is not well constrained by our approach. In this previous paper, some tests were performed to artificially increase the iron content (up to 10 wt%) but it was at the expense of fit quality (higher RMS value). Another possible explanation may be the choice of end-member for the dust analog. The addition of hematite in the mixture instead of palagonite, even if it does not provide as much acceptable fits (73% compared to 83% with the palagonite, Figure 9a), can lead to an increase of the Fe content of a few percent without changing the relative content of the other end-members. We cannot exclude that this potential artefact could come from the sensitivity of OMEGA measurements to chemically weathered surface compositions as explained below. One of the end-members that could account for more

Fe is olivine. The morphology and texture of the rocks observed in Gusev crater at the grain scale suggest that weathering on Mars should primarily affect the olivine during acidic episodes (McSween et al., 2006; Hurowitz et al., 2006). Resulting alteration rinds can then mask the olivine signature and thus bias the olivine detection. If this phenomenon is present at the global scale, it could then partly explain the lower Fe content derived from orbit with OMEGA compared from the GRS data that is a priori not affected by this bias. If this model is correct, the global scale OMEGA oxide compositions maps can bring context to those previously based *in situ* observations and the OMEGA-derived low Fe content cannot be interpreted in terms of igneous processes.

Magmatic trend (FeO+$Fe_2O_3$)/MgO vs $SiO_2$ can still be used as markers of dry or wet mantle sources (Figure 32b). The composition of all *in situ* measurements, of the majority of Martian meteorites and of GRS corresponds to tholeiitic basalts with lower $SiO_2$ content then the one obtained from OMEGA. Only the TES composition falls within the calc-alkaline field. However, this composition was suggested to be artifacts of alteration (McSween et al., 2009a) and thus deemed not representative of igneous processes. The OMEGA distribution overlaps the boundary between dry (TH on Fig. 32b) and wet (CA on Fig. 32b). The lower part of the distribution mostly corresponds to olivine enriched terrains with low Fe content (Table 4), while the upper part is attributed to regions where olivine is below the detection threshold (< 3 wt%). Those observations are in good agreement with the observed lack of Fe from the OMEGA-derived chemical composition compared to other measurements as discussed in the previous paragraph. A magma enriched in olivine is consistent with calc-alkali magma, however with the uncertainties on the OMEGA-derived chemical composition (on average 1.7 for the FeO*/MgO ratio) it is difficult to conclude whether this observation traces an actual difference in terms of water content in the magma.

For the first time oxide composition maps were derived at the km- and global scale on Mars. We show here that the surface may be globally altered for the few first microns of the surface leading to a poor Fe content compared to the one observed with GRS. The poor Fe content is correlated to the olivine and may suggest calc-alkali magma. We also predict almost no significant variations in terms of $SiO_2$ content (Figure 30) which shows that at this spatial sampling and with the end-member selected *a priori* in the mixture to model OMEGA reflectance spectra the dark regions of the Martian surface are almost uniquely of basaltic nature. These results significantly differ from in situ investigations indicating that Mars may have a far more varied near-surface crustal composition than suggested by orbital observations (e.g. Mangold et al. 2016).

## 6 – Conclusion

A radiative transfer model was implemented to process several millions of OMEGA spectra representative of igneous terrains of Mars. This task provided the modal composition and grain sizes at a planetary scale. The lithology can be summarized in five mineral maps at km-scale. Those compositional maps will be made available via the website PSUP (http://psup.ias.u-psud.fr). They give access to the distribution of pyroxenes (both low- and high-calcium pyroxenes), Mg-rich olivine, plagioclase and palagonite as a Martian dust analog. We excluded here all pixels that lack pyroxene signature, which notably included altered surfaces, most Amazonian terrains, and regions with hematite coverage. While the general compositions are not unexpected for Mars, their specific compositions and relative locations provide evidence for previously suspected and unexpected igneous compositional variability on its surface.

The pyroxene-bearing terrains that cover about 45 % of the surface between 60°S and 30°N are dominated by plagioclase with average content of ~50 vol%. This phase is most abundant in the Hesperian volcanic unit of Syrtis Major as well as in the Noachian terrains of Bosporos Planum and Protei Regio whose compositions differ from the rest of the Noachian areas. HCP is the second most abundant phase with average abundance of ~23 vol%. It is locally detected with abundance > 35 vol% in the volcanic Hesperian units of Syrtis Major and Hesperia Planum where it is associated with ejecta craters. LCP is widespread on the overall surface with average abundance close to 14 vol wt%. The highest LCP values are associated with highly cratered Noachian terrains of the southern hemisphere. At the global and km-scale, the LCP/(LCP+HCP) ratio tends to decrease with time. This observation is in good agreement with previous local and global studies (Mustard et al. (2005), Poulet et al. 2009b, Baratoux et al. 2013, Rogers & Hamilton 2015) and is interpreted to be the consequence of a cooling of the Martian mantle with time resulting into a decrease of the degree of partial melting with time. We bring here a new context and scale to this observation and show that it concerns the entire surface. Several tests were performed to assess the uncertainties on the derived composition and the choice of the dust analogue. Palagonite was selected to be the best dust analogue of the low albedo regions. This phase is relatively uniformly distributed on the surface and as expected it seems to be correlated with the albedo with average abundance close to 5 vol%. The olivine distribution differs from the rest of the end-members as its distribution is uneven with modelled abundance larger than the detection threshold (> 3 vol%) found only localized deposits. The largest abundance of olivine associated to pyroxene-bearing terrains is ~20 vol%. Those terrains are spatially well correlated with the olivine map based on spectral indexes.

A statistical analysis based on the *k-means* clustering routine was performed to classify the resulting abundances and highlight distinct mineralogical units at planetary scale. The modal mineralogy could be classified in 7 distinct mineralogical classes. The classes present various mineral assemblages with enrichment of at least one of the five components and thus highlight significant discrepancies in terms of composition. Apart from one class (#3, enriched in plagioclase) that accounts for more than one third of the Hesperian terrains modelled, this classification does not illustrate a compositional variation with time. We still observed the increase of HCP (class 5) at the expense of LCP (class 4) with time; however we also observed all types of mineral assemblages throughout the different sampled terrains. This suggests the presence of various and complex magmatic processes at a global scale during the Noachian and Hesperian and/or that in the Hesperian-aged terrains the footprints of previous magmatic episodes from the Noachian are still strongly present.

These new data enable us to evaluate the geochemistry of the crust as compositional elementary maps were derived from the modal composition maps. The OMEGA-derived composition is consistent with several distinctive geochemical characteristics previously considered as fingerprints of the Martian surface. It also supports the conclusion that the crust is basaltic and that no significant spread from this composition is observed at the km-scale. The major discrepancy with previous orbital investigations is the low value of Fe concentration obtained with OMEGA. Several explanations can be envisioned to account for this discrepancy between OMEGA and other investigations. The low value could translate shallow surface alteration highlighted here at the global scale, so that this geochemical trend cannot be interpreted in terms of igneous processes.

It is important to remind that the results of the modelling depend on the end-members selected beforehand. Various studies and tests were conducted to justify the choice of those end-members (Poulet and Erard (2004); Poulet et al. (2009a); Poulet et al. (2009b); this study). The numerous tests and comparisons performed as well as the uncertainties on each abundance (Table 1) provide a good confidence on the retrieved modal mineralogy. However, for the elemental composition, a wider selection of end-members could help better constrain some of the major components abundance (Si and Fe), that are restricted by the choice of end-members used in the modelling. The implementation of such tests will require new optical constants and important additional computation time. They could be envisioned in the future to reduce the uncertainties on the resulting chemical composition.

Finally, there is an aspect that has been almost totally eluded in this work: local variations at km-scale. Indeed, the major objective of this paper has been focused first on deriving maps, second on investigating global trend. But as pointed out in section 3, the mineral abundance maps reveal strong localized variations. Such variations have to be put in a geological context to better understand their origin. We believe that this approach will be very promising to provide constraints on the compositional heterogeneity of any mafic region/volcanic edifice and on mantle compositional models.

**Acknowledgments**

This work was partly supported by the French space agency CNES, CNRS and University de Paris-Saclay. This project has received funding from the European Union's Horizon 2020 (H2020-COMPET-2015) Research and Innovation Program under grant agreement 687302 (PTAL).